\documentclass[prd,amsfonts,twocolumn,nofootinbib,showpacs]{revtex4-1}
\usepackage{graphicx}% Include figure filestau
\usepackage{dcolumn}% Align table columns on decimal point
\usepackage{bm}% bold math
\usepackage{amsmath}
\usepackage{amsfonts} 
\usepackage{amssymb}
\usepackage{color}
\usepackage{aas_macros}
\usepackage{hyperref}
\usepackage{comment}
\usepackage{ mathrsfs }

\def\lnA{{\rm ln (10^{10} A_s) }}
\def\neff{{N_{\rm eff} }}

\newcommand{\ns}{n_{\rm s}}
\newcommand{\mTT}{C^{TT}_\ell}
\newcommand{\mTE}{C^{TE}_\ell}
\newcommand{\mEE}{C^{EE}_\ell}
\newcommand{\tTT}{\tilde{C}^{TT}_\ell}
\newcommand{\tTE}{\tilde{C}^{TE}_\ell}
\newcommand{\tEE}{\tilde{C}^{EE}_\ell}

\newcommand{\TT}{$\mTT$}
\newcommand{\TE}{$\mTE$}
\newcommand{\EE}{$\mEE$}
\newcommand{\CVL}{CVL}

\newcommand{\mfsky}{f_{\rm sky}}
\newcommand{\fsky}{$\mfsky$}

\newcommand{\LCDM}{$\Lambda$CDM}
\newcommand{\lmax}{\ifmmode \ell_{\mathrm{max}}\else
$\ell_{\mathrm{max}}$\fi}
\newcommand{\lmin}{\ifmmode \ell_{\mathrm{min}}\else
$\ell_{\mathrm{min}}$\fi}

\begin{document}
\title{CMB Polarization can constrain cosmology better than CMB temperature}

\author{Silvia Galli$^{a,b}$, Karim Benabed$^{a,b}$, Fran\c{c}ois Bouchet$^{a,b}$, Jean-Fran\c{c}ois Cardoso$^{a,d,e}$, Franz Elsner$^{f}$, Eric Hivon$^{a,b}$, Anna Mangilli$^{a,b}$, Simon Prunet$^g$, Benjamin Wandelt$^{a,b,c}$}

\affiliation{$^a$CNRS, UMR 7095, Institut d’Astrophysique de Paris, F-75014, Paris, 
France}
\affiliation{$^b$Sorbonne Universit{\'e}s, UPMC Univ Paris 06, UMR 7095, Institut d’Astrophysique de Paris,F-75014, Paris, France}
\affiliation{$^c$Department of Physics, University of Illinois at Urbana-Champaign, 1110 West Green Street, Urbana, Illinois, U.S.A.}
\affiliation{$^d$ Laboratoire Traitement et Communication de l’Information, CNRS
(UMR 5141) and T{\'e}l{\'e}com ParisTech, 46 rue Barrault F-75634
Paris Cedex 13, France}
\affiliation{$^e$ APC, AstroParticule et Cosmologie, Universit{\'e} Paris Diderot, 31 CNRS/IN2P3, CEA/lrfu, Observatoire de Paris, Sorbonne Paris
Cit{\'e}, 10, rue Alice Domon et L{\'e}onie Duquet, 75205 Paris Cedex 32 13, France}
\affiliation{$^f$ Department of Physics and Astronomy, University College London, London WC1E 6BT, U.K.}
\affiliation{$^g$Canada-France-Hawaii Telescope Corporation, 65-1238 Mamalahoa Hwy, Kamuela, HI 96743, USA}
  

\begin{abstract}

We demonstrate that for a cosmic variance limited experiment, CMB $E$ polarization 
{\it alone} places stronger constraints on cosmological parameters than CMB temperature.
For example, we show that \EE\ can constrain parameters better than \TT\ by up to a factor $2.8$ when a multipole range of $\ell=30-2500$ is considered.
We expose the physical effects at play behind this remarkable result and study how 
it depends on the  multipole range included in the analysis.  
In most relevant cases,  \TE\  or  \EE\ surpass the  \TT\  based cosmological constraints.
This result is important as the small scale astrophysical foregrounds are expected to have a much reduced impact on polarization, thus opening the possibility of building cleaner and more stringent 
constraints of the \LCDM\ model. This is relevant specially for proposed future CMB satellite missions, such as \textit{CORE} or \textit{PRISM}, that are designed to be cosmic variance limited in polarization  till very large multipoles.
We perform the same analysis for a Planck-like experiment, and conclude that even in this case \TE\  alone should determine the constraint on $\Omega_ch^2$ \emph{better} than \TT by $\sim 15\%$ , while determining
$\Omega_bh^2$, $n_s$ and $\theta$ with comparable accuracy.
Finally, we explore a few classical extensions of the 
\LCDM\ model and show again that CMB polarization alone provides more stringent 
constraints than CMB temperature in case of a cosmic variance limited experiment.
\end{abstract}


\maketitle
\section{Introduction}
\label{intro}
The results from the Planck satellite have recently confirmed that the cosmic 
microwave background (CMB) anisotropies are a powerful probe of cosmology \cite{planckcosmo}.
While these first cosmological results were based on Planck temperature data alone, 
interesting improvements are expected with the next release of data, when 
CMB polarization will be included in the analysis.
CMB polarization is often described in the literature as a unique source of information 
for reionization studies \cite{zaldarriaga97,mortonson08}, thanks to the 
large-scale signature that reionization is expected to leave in the polarization 
power spectra, and for inflation studies, as primordial gravitational waves are 
expected to produce B-mode CMB polarization \cite{seljak97,zaldarriagaspergel97,baumann09,bock09,kenney98,bicep}
and because polarization provides a cleaner probe of initial conditions 
\cite{spergel97,mortonson09,huspergel,durrer01,peiris03}. 
Furthermore, it is in general recognised that adding the information coming from 
polarization can improve the constraints on cosmological parameters and can help 
breaking some degeneracies \cite{seljak97,rocha03,colombo09,tegmark00,eisenstein99,bucher01,wu14,galli}.

In this paper, we argue that the CMB $E$ polarization data is much more than a mere improvement over 
the temperature anisotropies measurement. 
We demonstrate that either the temperature-polarization cross-correlation  \TE\  or
the  \EE\  polarization power spectra can provide tighter constraints on  
cosmological parameters than the temperature power spectrum \TT, in the case of a Cosmic Variance Limited  experiment (hereafter \CVL). 
The constraining power of \EE\ had already been noticed in \cite{rocha03}. Here
we show, for the first time, that \TE\ as well is more constraining than 
the \TT\ power spectrum, and explicit the physical reasons behind this conclusion.

We find that the \EE\ power spectrum can constrain the parameters (including the optical depth $\tau$) 
by up to a factor $2.8$ better than \TT\  in the case of a \CVL\ experiment probing up to multipoles $\ell=2500$, even when excising the 
large scales polarized signature of reionization ($\ell<30$). 
Overall, the  constraining power of the \EE\ power spectrum is found to be mildly dependent on the availability of
small scales ($\ell>2000$) and dramatically on the large scales ($30<\ell<130$). 
The \TE\  based constraints are also found to be tighter than the \TT\  based ones.

In the case of a more realistic Planck-like experiment, we demonstrate that the \TE\  
power spectrum provides comparable constraints to the \TT\  one, and a  
better one for the dark matter physical density $\Omega_{c}h^2$ by about $15\%$.

Finally, we show that even for classical extentions of the \LCDM\ models, the polarized 
based constraints are at least equivalent and often better than the temperature based one, 
for a \CVL\ experiment.

Our results open the possibility of 
improving the robustness of the CMB based cosmological constraints.
In fact, an important 
limitation of the \TT\  based cosmological constraints both from Planck 
\cite{2013arXiv1303.5075P} and from ground based experiments 
\cite{Dunkley:2013vt,Reichardt:2011il} is the presence of astrophysical foregrounds, particularly at  
small scales where the CMB temperature anisotropies are dominated by the 
contribution from unresolved radio and infrared galaxies.
Polarization, however, is expected to suffer less from this contamination \cite{seiffert07,tucci04}.
Using the exceptional constraining power of the polarized
CMB observations, one should thus be able to build cross-checks of the temperature
results and improve the temperature foreground models at small scales.

In Section \ref{sec:fisher}, we first describe the Fisher Matrix 
formalism that will be used to calculate forecasts on cosmological parameters. Then,
in Section \ref{sec:CVL} we present the main results on \LCDM\  parameters for a \CVL\ 
experiment. In particular, we show how the constraints depend 
on the maximum or minimum multipole included in the analysis, and describe the physics 
at play in this setting. Section \ref{sec:planck} reproduces 
the analysis for a Planck-like experiment. The sensitivity of the constraints 
to prior knowledge of the reionization optical depth $\tau$ is discussed in  Section 
\ref{sec:tau}, and we show how the large scales ($\ell<30$) contribute to the determination of $\tau$.
Finally, Section \ref{sec:exten} allows us to generalize our conclusions to classical \LCDM\  model extensions.

\section{Methodology}
\label{sec:fisher}
\begin{figure}[thb]
\centering
 \includegraphics[angle=0,width=0.4\textwidth]{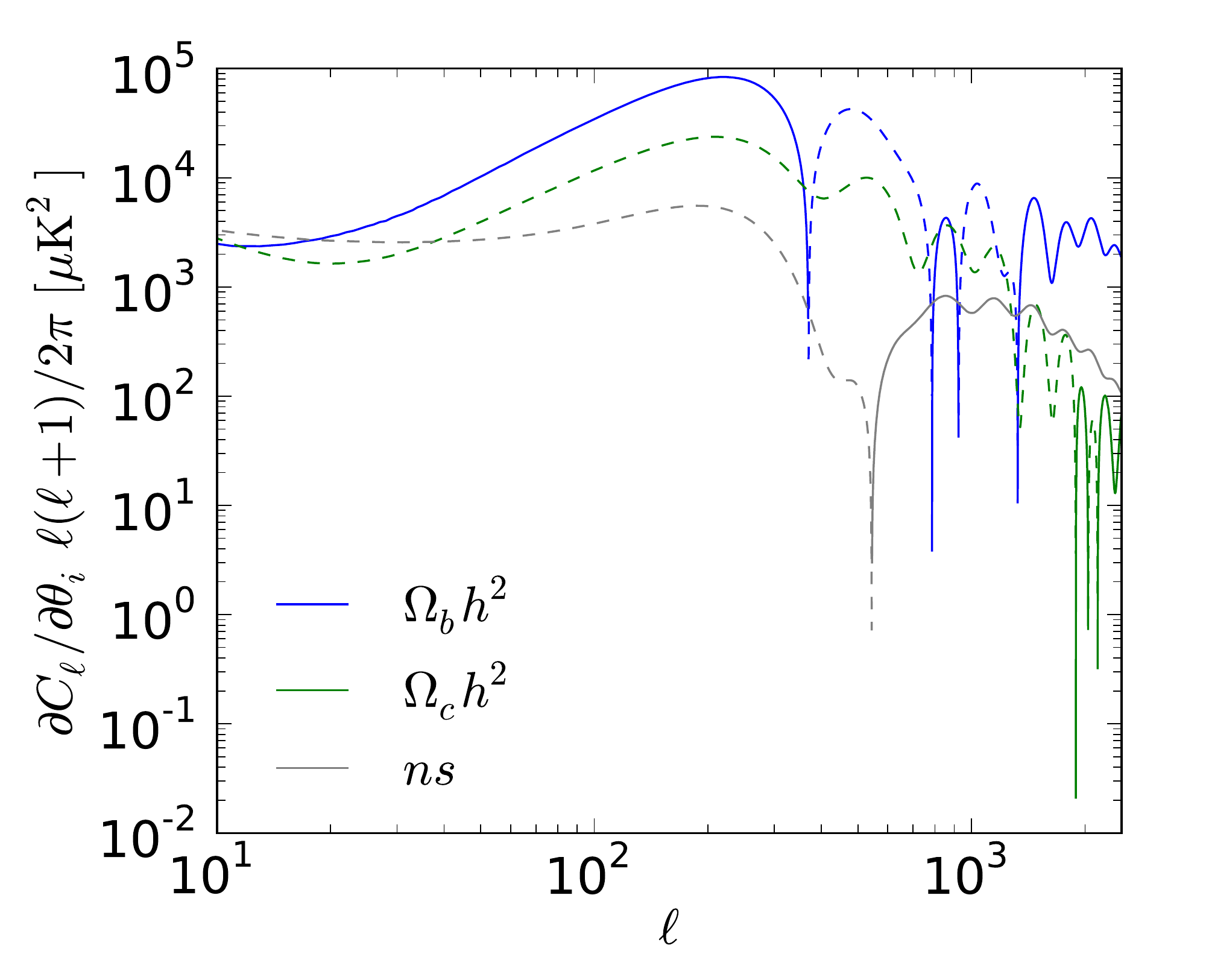}
 \includegraphics[angle=0,width=0.4\textwidth]{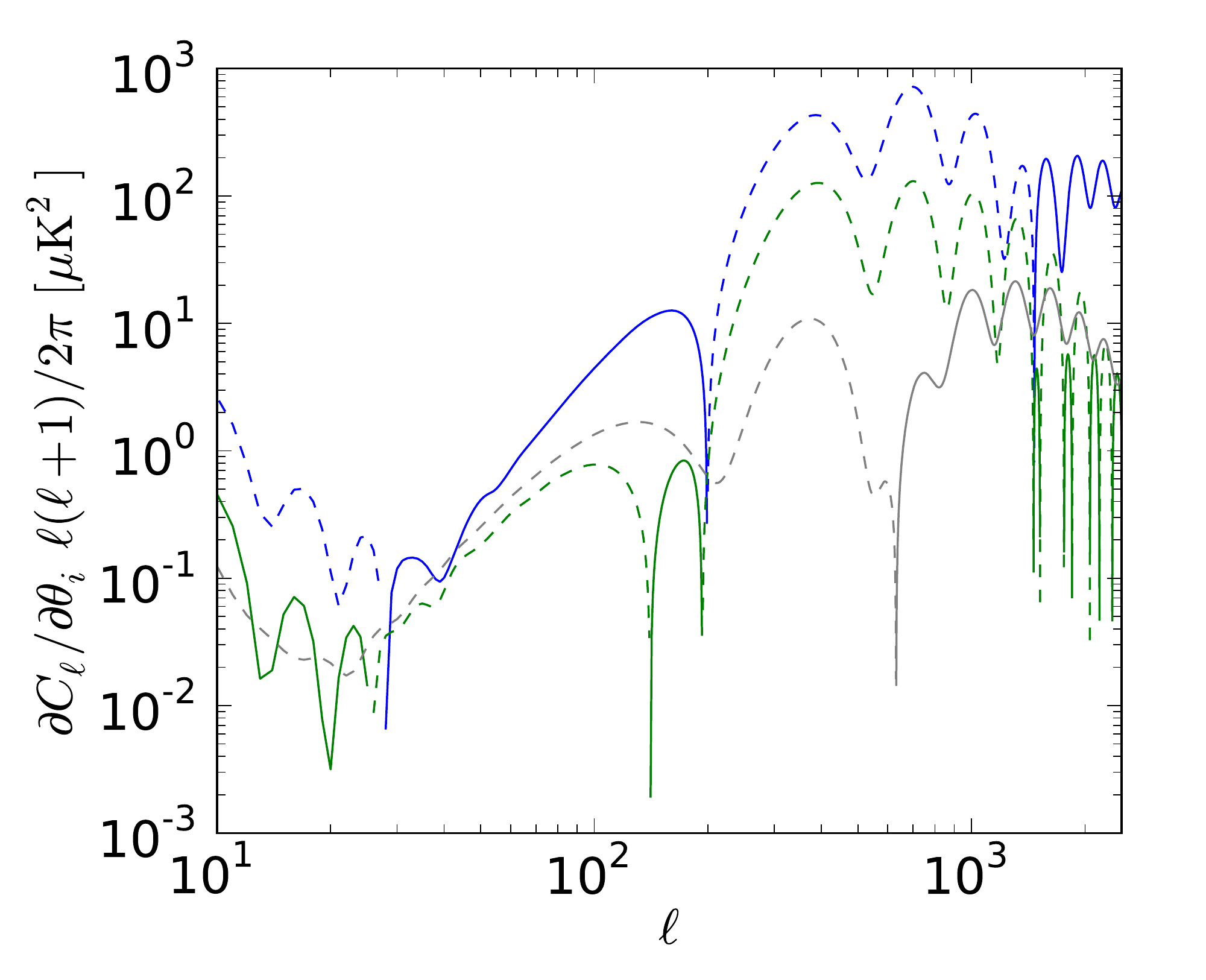}
\includegraphics[angle=0,width=0.4\textwidth]{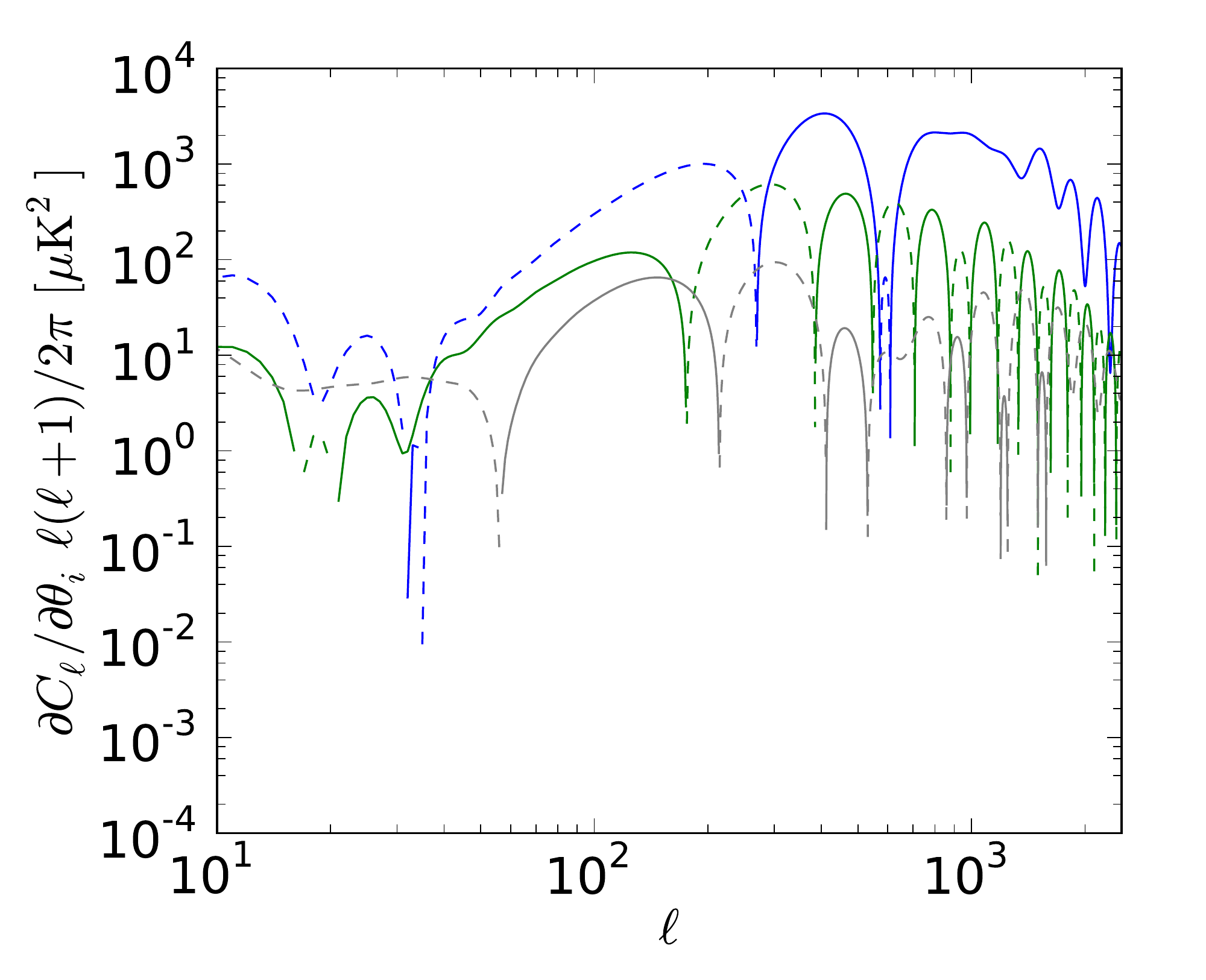}
 \caption{Derivatives of the  \TT (top panel),  \EE (middle panel) and  \TE (bottom panel) power spectrum with respect to the $\Lambda$CDM parameters. The plot is in logarithmic scale, so we plot the absolute value of the derivatives, showing negative values as dashed lines.}%%
\label{derivativesEE}%%
\end{figure}%%

\begin{figure}[t]
\centering
\includegraphics[angle=0,width=0.4\textwidth]{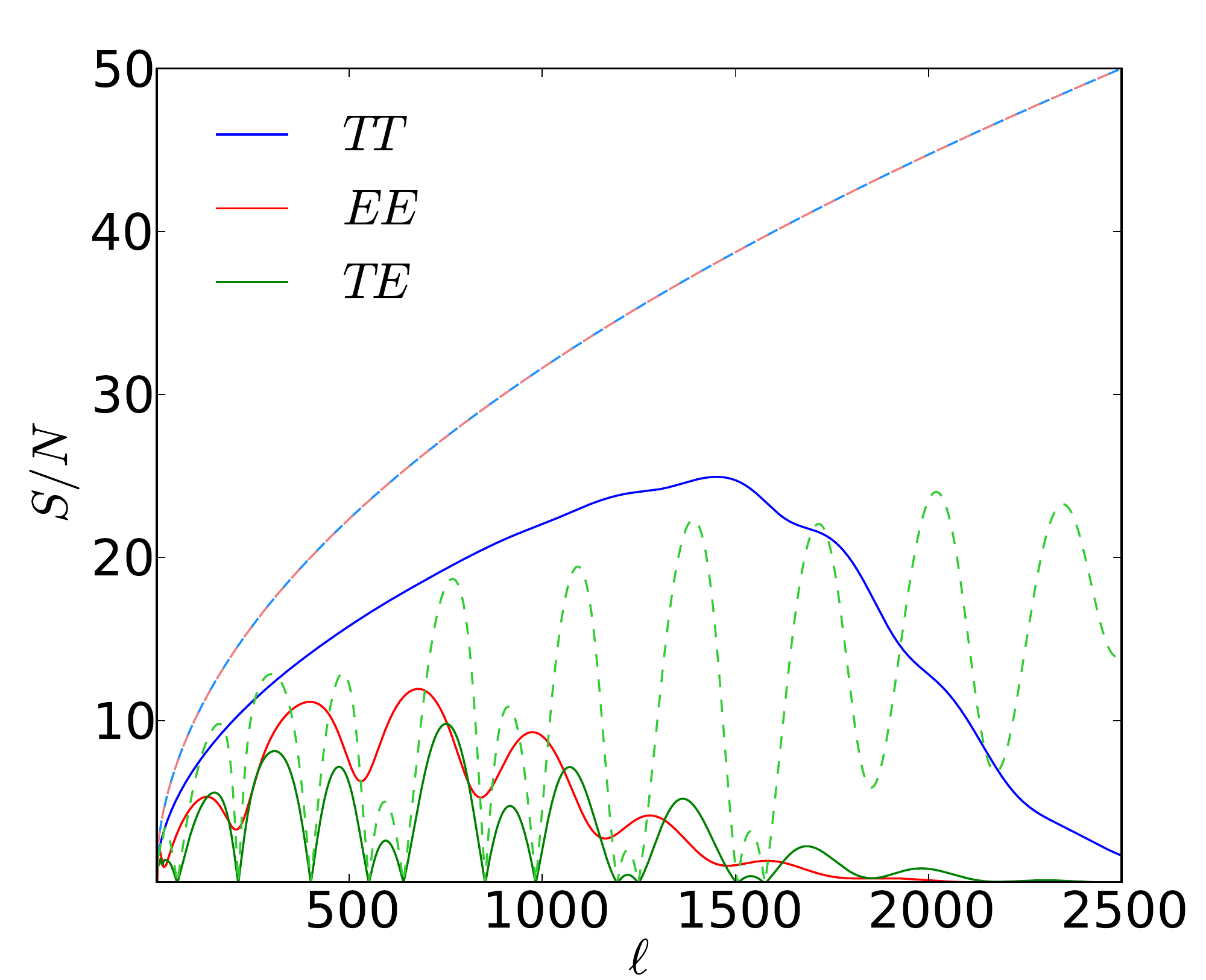}

\caption{ Signal-to-noise for a Planck-like full mission experiment (solid lines) with $f_{\rm sky}=0.5$, as detailed in Table \ref{tab:exp},  for the  \TT\  (blue),  \EE\ (red) and  \TE\  (green) power spectra. Dashed lines show the signal to noise for a \CVL\  experiment with $f_{\rm sky}=1$.}

\label{spectra}
\end{figure}

The information on a vector of cosmological parameter $\theta_i$ that can be 
extracted from a subset $M$ of the CMB temperature and polarization power spectra
$(\mTT,\mTE,\mEE)$ can be estimated using the Fisher Information Matrix (FIM) 
(see e.g. \cite{tegmark97,Jungman96}):
\begin{equation}
 F^{\rm CMB}_{ij}=\sum_{X,Y\ \mathrm{in}\ M}\sum_{\ell} \frac{\partial {C}_\ell^{X} }{\partial\theta_i} [\mathscr{C}_\ell]^{-1}_{XY} \frac{\partial C_\ell^{Y} }{\partial\theta_j}.
\label{fishercmb} 
\end{equation}
One can forecast the cosmological constraints obtained using either \TT, \TE, \EE\ or 
all of them by changing the content of $M$ in the above equation.
To compute the elements of the FIM, we need to evaluate the power spectra covariance matrix
and the derivatives of the power spectra with respect to the parameters of interest.

The computation of the covariance matrix can be simplified using the following assumptions.
We ignore all sources of non-Gaussianity in the CMB data.
We model the lensing of the CMB by the large scale structure as a smoothing of the 
observed CMB power spectra, and ignore the lensing induced 4-point correction which is being known to 
introduce a small correlation in the \EE\ covariance matrix and a negligible one on \TT\  
\cite{2012PhRvD..86l3008B}. We further assume that foreground contamination, beam
uncertainties and other systematics have been corrected to a level much smaller 
than the statistical error and  we do not  take into account any marginalization over their parameters. 
Finally, we account for partial sky coverage (\fsky) by a rescaling of the full 
sky covariance and ignore the mode correlations it introduces.
Under those assumptions, the covariance matrix $\mathscr{C}_\ell$ is only a function of
the theoretical CMB power spectra, and of the effective noise power spectra of 
temperature ($N^{TT}_\ell$) and $E$ polarization ($N^{EE}_\ell$). We have: 
\begin{widetext}

\begin{equation}
\mathscr{C}_{\ell}=\frac{2}{2 \ell+1}\frac{1}{\mfsky}\left( 
\begin{array}{ccc}
(\tTT)^2      & (\tTE)^2      & \tTT\tTE                \\
(\tTE)^2      & (\tEE)^2      & \tEE \tTE               \\
\tTT \tTE\,\, & \tEE \tTE\,\, & 1/2[(\tTE)^2+\tTT \tEE] \\
\end{array} 
\right)
\end{equation}
\end{widetext}

with $\tTT$, $\tEE$\ and $\tTE$, the observed temperature and polarization spectra, given by
\begin{eqnarray}
\tTT &=& \mTT + N^{TT}_\ell, \\
\tEE &=& \mEE + N^{EE}_\ell, \\
\tTE &=& \mTE
\end{eqnarray} 

The numerical evaluation of the power spectra and their derivatives will be performed using 
the 
\texttt{PICO}\footnote{https://sites.google.com/a/ucdavis.edu/pico/home} 
code \cite{picopaper}. An example of  some of the power spectra derivatives 
is presented in Fig. \ref{derivativesEE}.

In the next sections, we forecast the constraints on the parameters of a \LCDM\  cosmology.
We focus on the following $6$ parameters: the physical baryon and CDM densities,
$\omega_b=\Omega_bh^2$ and $\omega_c=\Omega_ch^2$, the angular dimension of the 
sound horizon at recombination $\theta$, the normalization of the primordial 
power spectrum, $\lnA$, with pivot scale $k_0=0.05 \,\mathrm{Mpc^{-1}}$, the scalar 
spectral index, $n_{s}$,  and the optical depth to reionization, $\tau$. We also 
explore a few classical extentions of the model in Section \ref{sec:exten}.
The fiducial values of the parameters used thorough the article are presented in 
Table \ref{tab:fiducial}.

We consider two experimental settings: a full sky cosmic variance limited experiment 
(CVL) and a Planck-like mission.
For the CVL experiment, we have $N_\ell^{TT}=N_\ell^{EE}=0$ and $\mfsky=1$. 
Changing the smallest available scale ($\lmax$) as we do in section \ref{sec:CVL} allows the reader to 
evaluate quickly what will be the behaviour of a more realistic experiment.

The Planck-like mission corresponds to $30$ months worth of
combined observations by the two best Planck CMB channels at $\rm 143 \, GHz$ and $\rm 217 \, GHz$. 
We assume that due to galactic dust contamination only half of the sky is 
available for cosmological use, $\mfsky=0.5$.
The effective noise of the combined data is given by \cite{knox95}
\begin{eqnarray}
\label{noisecmb}
N^{X,chan}_{\ell} &=& \left(\frac{w_{X,chan}^{-1/2}}{\mu{\rm K\mbox{-}rad}}\right)^2 
\exp\left[\frac{\ell(\ell+1)(\theta_{\rm pix}/{\rm rad})^2}{8\ln 2}\right],\\
N^{X}_{\ell}&=&\frac{1}{\sum_{chan}\{1/(N^{X,chan}_{\ell})\}}
\end{eqnarray}
where $N^{X,chan}_{\ell}$ is the noise in each of the channels considered in temperature $(X=T)$ or polarization $(X=P)$, while 
$N^{X}_{\ell}$ is the total noise, $w_X^{-1}=\{(\Delta X/T)\times T_{CMB}\times\theta_{\rm pix}\}^2[\mu K^2]$ is the raw sensitivity,  $(\Delta X/T_{CMB})$ 
is the sensitivity per pixel, $T_{CMB}=2.725K$ is 
the $CMB$ temperature, and  $\theta_{\rm FWHM}$ is the full width at half maximum 
beam size. We compute the characteristics of this experiment according to the 
Planck Blue Book \cite{blubook}, and report the numerical values in Table \ref{tab:exp}. 
This setting gives a reasonable estimate of the expected sensitivity of the
second Planck data release. 
We show the signal-to-noise ratio (including cosmic variance) for different power spectra both for a \CVL\ and Planck-like experiment in Fig. \ref{spectra}.

\begin{table}
\begin{center}
\begin{tabular}{ |l |l |  }
\hline \hline
Parameter &Fiducial\\\hline
$\Omega_bh^2$&$0.022032$\\
$\Omega_ch^2$&$0.12038$\\
$\theta100$&$1.04119$\\
$\tau$&$0.0925$\\
$n_s$&$0.9619$\\
$\lnA$&$3.0980$\\ 
$\Omega_k$&$0$\\
$Y_p$&$0.24$\\
$N_{eff}$&$3.046$\\
$n_{run}$&$0$\\
$\Sigma m_\nu$&$0$\\
\hline
\hline
\end{tabular}
  \caption{Fiducial model used in the analysis.}
\label{tab:fiducial}
\end{center}
\end{table}

\begin{table}[!htb]
\begin{center}%%
\begin{tabular}{cccc}%%
Channel & FWHM & $\Delta T/T$ & $\Delta P/T$ \\
GHz & arcmin & $[\mu K/K]$ & $[\mu K/K]$ \\
\hline
%Planck 2 channels & 143 & 7'& 2.2 & 4.4\\
%Nom $f_{\rm sky}=0.5$ &217 & 5'& 4.8 & 9.6\\ \hline
143 & 7'& 1.5 & 3.0\\
217 & 5'& 3.3 & 6.6\\ \hline
\end{tabular}
\caption{ Specifications of the Planck-like experiment. Values are based on the Blue Book specification and for a 30 month mission \cite{blubook}.
 $\Delta T/T$ is the sensitivity per pixel.}
\label{tab:exp}
\end{center}
\end{table}

\section{Cosmic variance limited experiment}
\label{sec:CVL}
We present in this section forecasts on   cosmological parameters in case of a \CVL\  Experiment for  \TT,  \EE\ or  \TE\  considered separately or from the combination of all the spectra.
 Our baseline in this Section has $\lmin=30$ and $\lmax=2500$ unless otherwise specified, and always includes a prior\footnote{The prior we use is comparable with the constraint obtained by the WMAP satellite observing polarization at large scales \cite{wmap9}.} on $\tau$, $\sigma(\tau)=0.013$. In this way, we can compare the information content of the different power spectra assuming  nearly the same information about reionization, that can  otherwise be more tightly constrained observing its signature at large scales ($\ell\lesssim 30$) in polarization. We will discuss the effect of relaxing this $\tau-$prior assumption in Section \ref{sec:tau}.

\subsection{Evolution of the constraints with $\lmax$}

\begin{figure*}[t]
\centering
 \includegraphics[angle=0,width=0.7\textwidth]{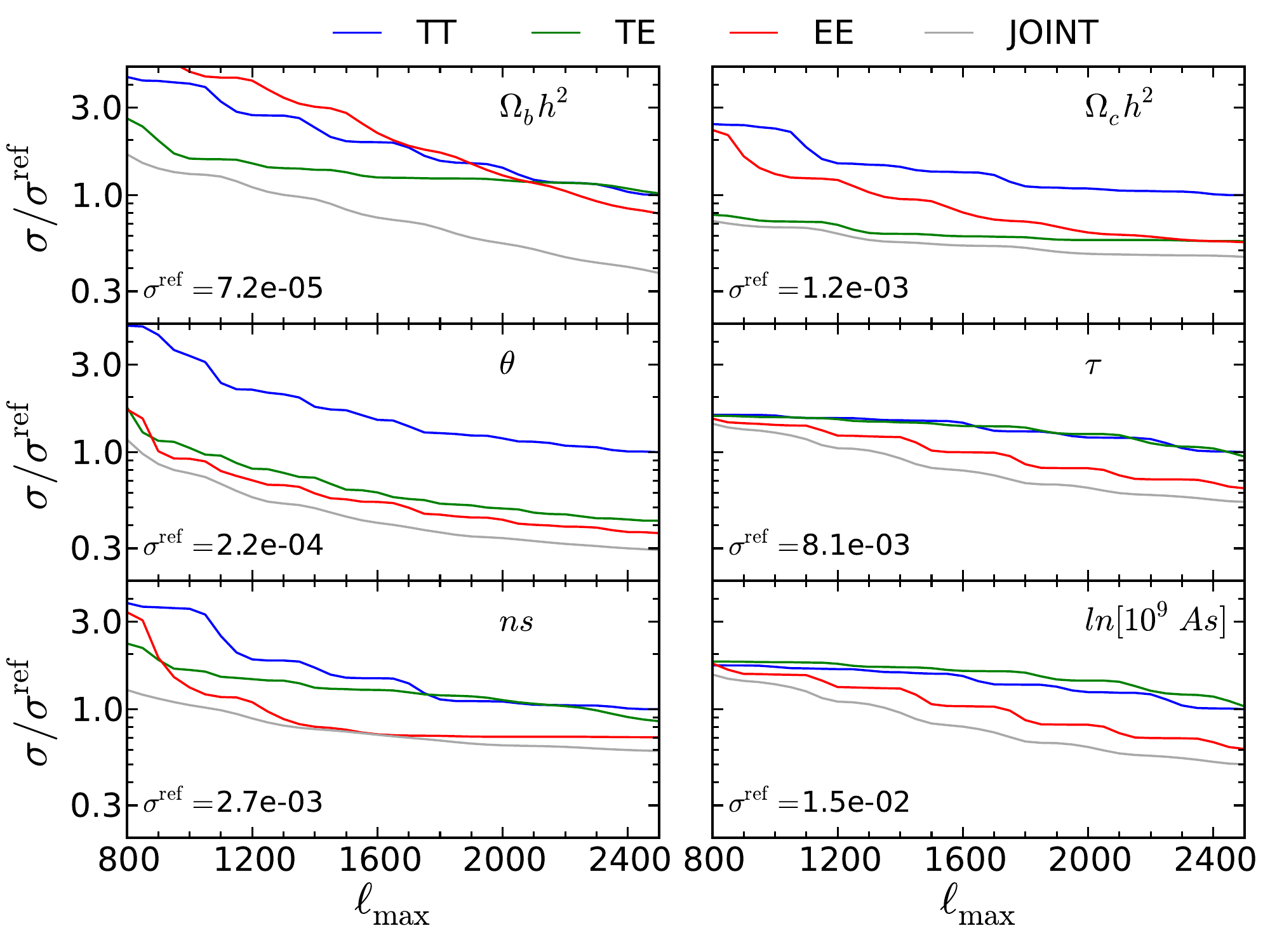}
\caption{Standard deviations on $\Lambda CDM$ parameters as function of $\lmax$, normalized to the standard deviation $\sigma^{\rm ref}$ obtained from \TT\ with $\lmax=2500$. We consider a \CVL\ experiment with $\lmin=30$ and a prior on $\tau$. We consider here lensed CMB power spectra. }
\label{lensedCVL}
\end{figure*}
\begin{figure}[t]
\centering
 \includegraphics[angle=0,width=0.5\textwidth]{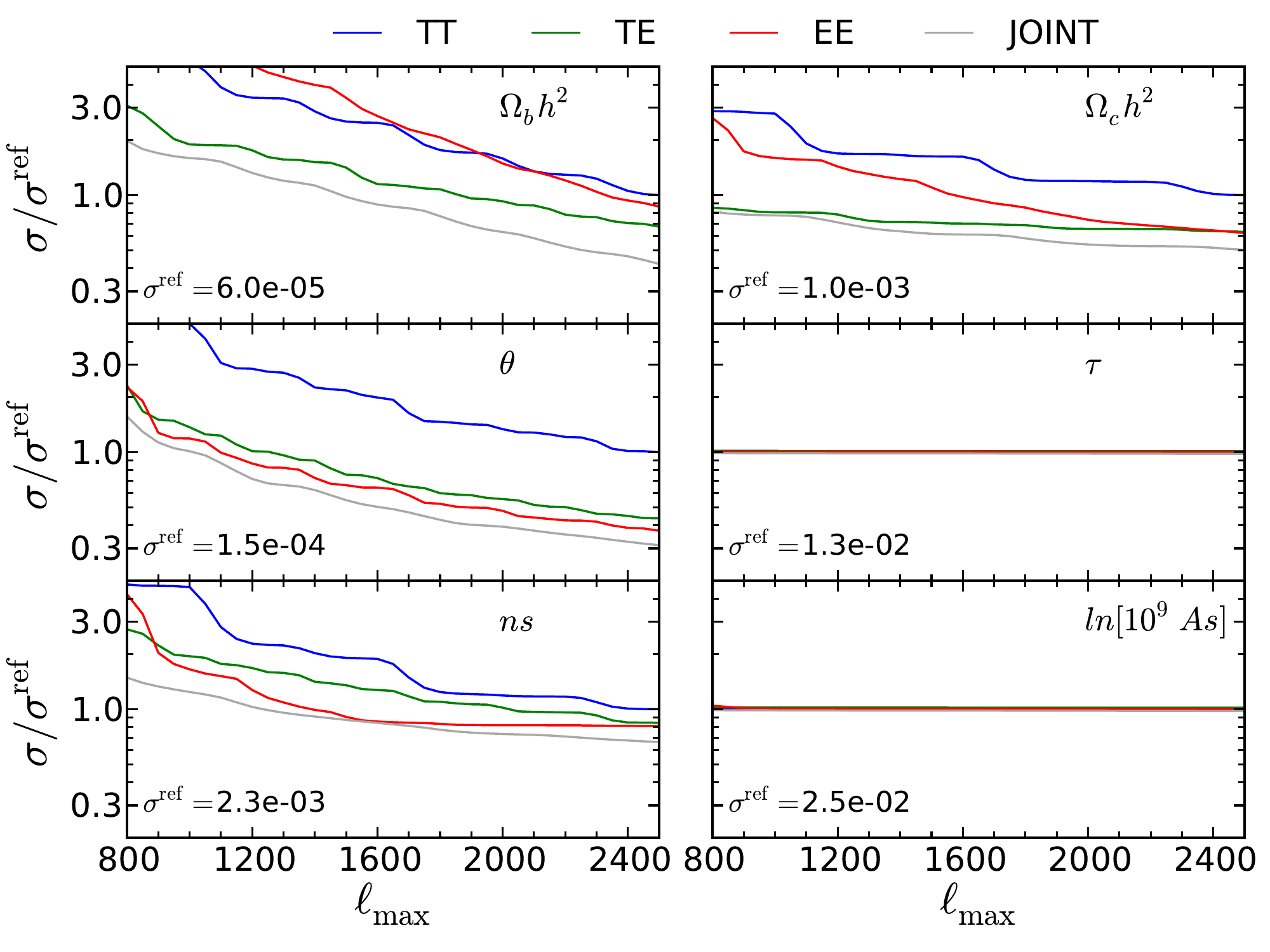}
\caption{ Same as Fig. \ref{lensedCVL}, but for unlensed CMB power spectra. }
\label{scalarCVL}
\end{figure}

\begin{figure}[t]
\centering
 \includegraphics[angle=0,width=0.5\textwidth]{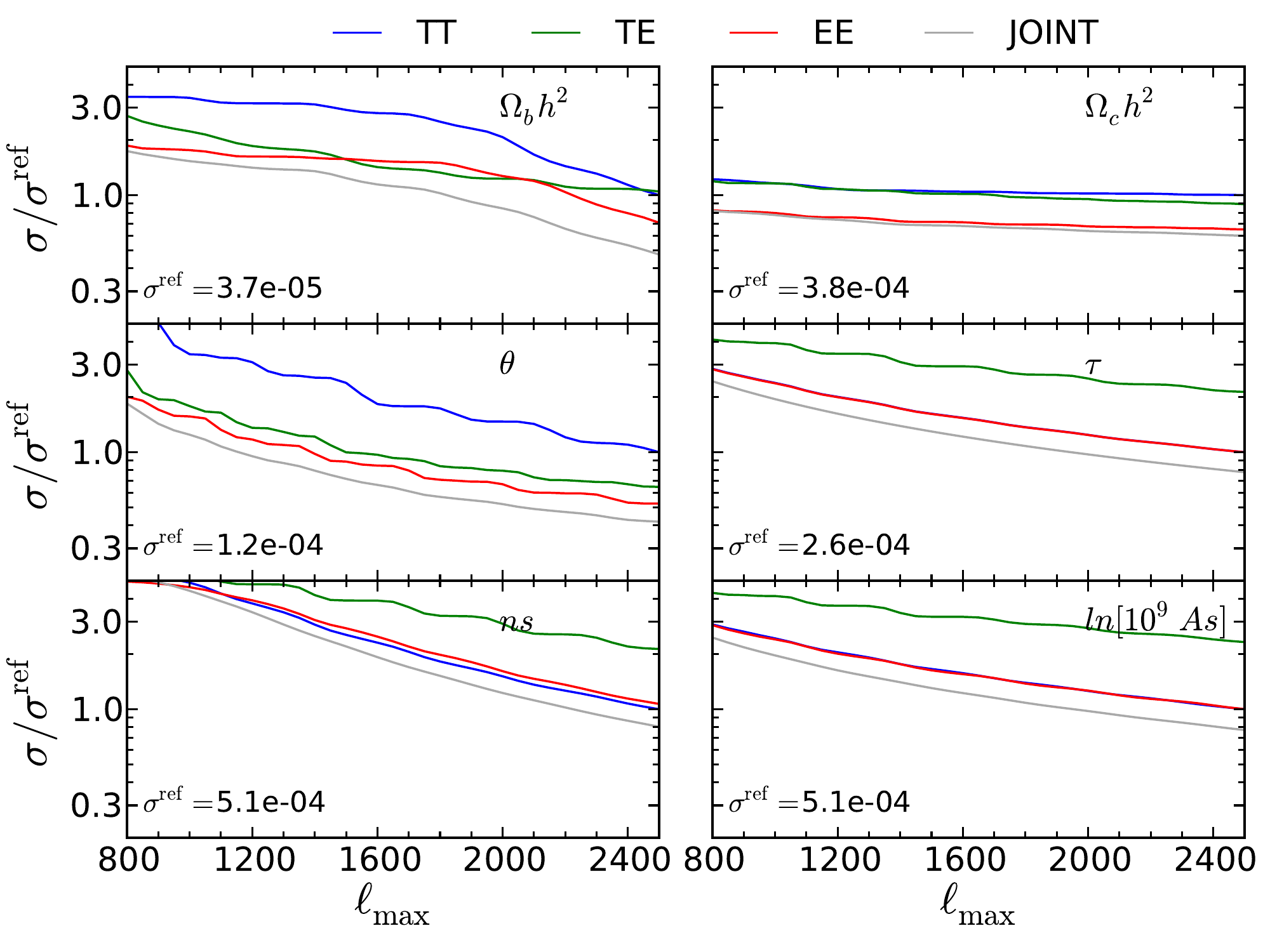}
\caption{Same as Fig. \ref{lensedCVL}, but the constraints in this case are calculated as the inverse of the diagonal of the Fisher Matrix, i.e. they are not marginalized over degeneracies between parameters. By comparing with Fig. \ref{lensedCVL}, one can determine whether constraints on the parameters are limited by their degeneracies or by the sensitivity of the spectra to each parameter separately.}
\label{diaglensedCVL}
\end{figure}

In this section we analyze how the constraints change as a function of maximum multipole ($\lmax$) included.
In Fig. \ref{lensedCVL} we show how the constraints evolve with $\lmax$ for lensed CMB spectra (reference constraints).
By observing Fig. \ref{lensedCVL}, we notice that at $\lmax=2500$,  \EE\ is the best at constraining all parameters, as also shown in Table \ref{tab:lcdm} and as already noticed by \cite{rocha03}. 

\begin{table*}[thb]%\footnotesize
\begin{center}
\begin{tabular}{rcccccccccccc}
\hline\hline
\noalign{\vskip 3pt}
Data& $\Omega_bh^2$ & & $\Omega_ch^2$ & & $\theta$ & & $\tau$ & & $\ns$ & & $\lnA$ \\
\noalign{\vskip 3pt}
\hline
\noalign{\vskip 3pt}

 \TT  & $7.2\times 10^{-5}$ & [0.3\%] & $1.2\times 10^{-3}$ & [1.0\%] & $2.2\times 10^{-4}$ & [0.02\%] & $8.1\times 10^{-3}$ & [9\%] & $2.7\times 10^{-3}$ & [0.3\%] & $1.5\times 10^{-2}$ & [0.5\%] \\
 \EE  & $5.7\times 10^{-5}$ & (1.3)   & $6.5\times 10^{-4}$ & (1.9)   & $7.9\times 10^{-5}$ & (2.8)    & $5.2\times 10^{-3}$ & (1.6) & $1.9\times 10^{-3}$ & (1.4)   & $9.0\times 10^{-3}$ & (1.7)   \\
 \TE  & $7.3\times 10^{-5}$ & (1.0)   & $6.6\times 10^{-4}$ & (1.8)   & $9.3\times 10^{-5}$ & (2.4)    & $7.7\times 10^{-3}$ & (1.1) & $2.3\times 10^{-3}$ & (1.2)   & $1.5\times 10^{-2}$ & (1.0)  \\
JOINT & $2.7\times 10^{-5}$ & (2.7)   & $5.4\times 10^{-4}$ & (2.2)   & $6.4\times 10^{-5}$ & (3.4)    & $4.4\times 10^{-3}$ & (1.8) & $1.6\times 10^{-3}$ & (1.7)   & $7.4\times 10^{-3}$ & (2.0)  \\
\noalign{\vskip 3pt}
\hline

\end{tabular}

\caption{Standard deviations on the $\Lambda$CDM model for a \CVL\  experiment from  \TT,  \EE,  \TE\  taken separately or from the combination of all the power spectra. These constraints are calculated assuming $\lmin=30$, $\lmax=2500$, and a prior on $\tau$, $\sigma(\tau)=0.013$. In square brackets, on the first line we translate the standard deviation in relative error. In parenthesis, on the next lines we show the improvement factor compared to the  \TT\  case.}

\label{tab:lcdm}
\end{center}
\end{table*}
In particular, the  \EE\ power spectrum constrains the angular size of the sound horizon $\theta$ better than  \TT\  by a factor $\sim 2.8$, $\Omega_ch^2$, $\lnA$, $\tau$, $\ns$  by  a factor $1.9$, $1.7$, $1.6$ and $1.4$ respectively, while the smallest improvement is on the baryon density $\Omega_bh^2$ by a factor $\sim 1.3$.

The  \TE\  power spectrum constrains the dark matter density $\Omega_ch^2$ as strongly as  \EE, $\theta$ and $\ns$ slightly worse than  \EE\ (only by a factor $\sim 1.2$), while it constrains the other parameters ($\Omega_bh^2$, $\lnA$ and $\tau$) at a level comparable to  \TT. It is interesting to notice however that  \TE\  becomes the best at constraining the matter densities $\Omega_bh^2$ and $\Omega_ch^2$ if $\lmax\lesssim 2100$, while the other parameters keep being constrained best by  \EE.

Finally, combining all the spectra leads to a substantial improvement only to  the constraint on $\Omega_bh^2$ (by a factor $\sim 2.1$ compared to  \EE\ alone), while the other parameters  improve only by a factor $\sim 1.2$ compared to the constraints from  \EE\ alone.

In order to gain some physical insight into why the constraints are stronger when determined from the  \EE\ power spectrum, we will make use of two additional figures.  Fig. \ref{diaglensedCVL}  shows the constraints obtained as the inverse of the diagonal of the Fisher Matrix, i.e. { \it without marginalizing} over the degeneracies among all the parameters (we will refer to this as the `diagonal' case). This is equivalent to calculating constraints for one parameter at the time, assuming the others fixed to their fiducial value. This plot is useful to understand whether the reference constraints in Fig. \ref{lensedCVL} are limited by degeneracies among parameters or by the intrinsic information encoded in the CMB, limited only by cosmic variance. Fig. \ref{scalarCVL} shows the same plots as Fig. \ref{lensedCVL} but for unlensed CMB spectra. This plot is useful to assess whether the information coming from the  effect of weak lensing at high multipoles impacts the constraints on cosmological parameters. We also report in Table \ref{lensedscalarratio} the ratio of the constraints obtained from the scalar versus lensed power spectra.

\begin{table}[thb]%\footnotesize
\begin{center}
\begin{tabular}{lcccccc}
\hline\hline
\noalign{\vskip 3pt}
Data & $\Omega_bh^2$ &  $\Omega_ch^2$ & $\theta$  & $\tau$  & $\ns$  & $\lnA$\\ 
\noalign{\vskip 3pt}
\hline
\noalign{\vskip 3pt}
\multicolumn{7}{l}{\CVL}\\
\noalign{\vskip 3pt}
\hline
\noalign{\vskip 3pt}
TT&$0.8$&$0.9$	&$0.7$&	$1.6$&	$0.8$&	$1.7$\\
EE&$0.9$	&$1.0$&	$0.7$&	$2.5$&	$1.0$&	$2.8$\\
TE&$0.6$	&$1.0$&	$0.7$&	$1.7$&	$0.8$&	$1.7$\\
JOINT&$0.9$&	$1.0$&	$0.8$&	$2.9$&	$0.9$&	$3.3$\\
\noalign{\vskip 3pt}

\hline
\noalign{\vskip 3pt}
\multicolumn{7}{l}{Planck}\\
\noalign{\vskip 3pt}

\hline
\noalign{\vskip 3pt}

TT&1.0&	1.0&	0.9&	1.1&	1.0&	1.2\\
EE&1.1	&1.1	&0.9&	1.0&	1.1&	1.0\\
TE&1.0	&1.1	&0.9&1.0&	1.1&	1.0\\
JOINT&1.0&	1.1&	0.9&	1.2&	1.0&	1.2\\
\noalign{\vskip 3pt}
\hline
\end{tabular}

\caption{Ratio of standard deviations obtained from unlensed power spectra divided by the standard deviations obtained from lensed power spectra. We show results for  \TT\,  \EE\,  \TE\ power spectra taken separately or from the joint combination of all of them, for a \CVL\ and a Planck-like full mission experiment. These constraints are calculated assuming $\lmin=30$, $\lmax=2500$, and a prior on $\tau$, $\sigma(\tau)=0.013$. This table shows that the lensed spectra do not always provide the tightest constraints.}
\label{lensedscalarratio}
\end{center}
\end{table}

 By comparing these plots, we notice the following features:
\smallskip

\noindent\textbf{Constraint on $\theta$.}
As already mentioned and shown in Fig. \ref{lensedCVL}, the angular size of the sound horizon $\theta$ is better determined using the  \EE\ or  \TE\  power spectra alone than by  using  \TT\  alone.
As is well known, the effect  of changing  $\theta$ on the power spectra is to shift the position of the acoustic peaks, e.g. an increase in $\theta$ shifts the peaks to larger scales (smaller multipoles). However, peaks in  \EE\ and  \TE\  are \emph{sharper} than in  \TT\  \cite{zaldarriaga04}, so that their position can be better determined from polarization. This is due to the fact that polarization is mainly sourced by the gradient of the velocity field of the photo-baryonic fluid, while  temperature is sourced both by the perturbations in the density field (Sachs-Wolfe plus intrinsic temperature effects) and by the perturbations in the velocity field (Doppler effect). The peaks in polarization are thus only determined by the extrema of oscillations in the velocity field, while the peaks in  temperature are mainly determined by the oscillation extrema of the density distribution plus the smaller contribution (damped by the presence of baryons that slow down the sound speed) from oscillations of the velocity field.  The peaks in temperature, determined by these two effects  out of phase by $\pi/2$, are thus smoother compared to the ones in polarization, which can therefore better constrain $\theta$.
Furthermore, by comparing Fig. \ref{lensedCVL} with the diagonal case in Fig.  \ref{diaglensedCVL}, we notice that the constraints on $\theta$ are only mildy affected by degeneracies with other parameters, as the errorbars obtained in the diagonal case are only a factor $\sim 2$ smaller than the ones obtained with the full marginalization over all parameters.
Finally, by comparing Fig. \ref{lensedCVL} with the diagonal case in Fig. \ref{diaglensedCVL}, we notice that $\theta$ would be better constrained by the unlensed CMB spectra, due to the fact that lensing smooths the peaks/throughs of the CMB power spectra, making the position of the peaks harder to determine. 
% It is also interesting to notice that even in the diagonal case  \TE can constrain $\theta$ at comparable level to  \EE. 
\smallskip

\noindent\textbf{Breaking the $\tau-\lnA$ degeneracy with lensing.}
The  \EE\  power spectrum alone is also better at constraining the reionization optical depth and the amplitude of the primordial power spectrum $\lnA$.
One of the main effects of increasing the reionization optical depth $\tau$ on the power spectra is to suppress the amplitude of the peaks at scales smaller than the causal horizon at the epoch of reionization. This effect is due to the fact that only a fraction $\exp{(-\tau)}$ of the photons manage to stream freely through the reionized universe, so that the power spectra are in fact proportional to $\rm \exp{(-2\tau)}$ \cite{zaldarriaga97,holder03}. This effect is highly degenerate with the amplitude of the primordial power spectrum $A_{\rm s}$ \cite{bond87}.
The degeneracy can be lifted in two ways. Either measuring the reionization bump in the polarization power spectra at large scales ($\ell\lesssim 30$), that  directly constrain $\tau$, or measuring with high accuracy the effect of lensing on small scales ($\ell\gtrsim 1000$), as the amplitude of the lensing potential depends on the amplitude of the initial perturbations $A_{\mathrm s}$ but not on $\tau$. 
In the cases we consider in this Section, we do not include the large scale polarization in our calculations, but we weakly break the degeneracy between the two parameters by adding a prior on $\tau$ with width $\sigma(\tau)=0.013$.
From Fig. \ref{lensedCVL}, we notice that the constraints on $\tau$ and $\lnA$ improve when including higher $\lmax$, thanks to the degeneracy-breaking effect of lensing. Such an improvement is not present when the same constraints are evaluated from unlensed power spectra, as shown in Fig. \ref{scalarCVL}, where the constraints on $\tau$ and $\lnA$ are  constant with $\lmax$.  Furthermore, we notice from the reference constraints in Fig. \ref{lensedCVL}  that \EE\ provides a better measurement than  \TE\  or  \TT\  on these two parameters. This is due to the fact that the effect of lensing is larger on polarization than on temperature (at $\ell\sim 2000$, the change in the amplitude is a factor $2$ larger for \EE\ than \TT) \cite{zaldarriaga98}. A change in the lensing amplitude thus impacts  polarization more strongly than temperature, resulting in a stronger constraint on $\lnA$. This is also clear from Table \ref{lensedscalarratio},  showing that the lensed \EE\ power spectrum  constrains $\tau$ by  a factor $3$ better than unlensed spectra\footnote{We remind here that both in the lensed and unlensed case we include a prior on $\tau$ in the analysis.}, while the constraints from the lensed \TE\ and \TT\ improve only by a factor $1.7$.

\smallskip

\noindent\textbf{A general remark on  \TE}.
To understand why \EE\ determines $\Omega_bh^2$, $\Omega_ch^2$ and $\ns$ as well as or better than  \TT, we refer the reader to the next Subsection \ref{sectionlmin}. 

As a general remark, however, we would like to emphasize that  \TE\  is very powerful at constraining parameters despite the fact that its signal-to-noise ratio (both in the \CVL\  and in the Planck case) is not as large as the one for  \TT\  or  \EE\ alone, as shown in Fig  \ref{spectra}. This is however not surprising, as the determining factor for strong constraints on the parameters is not only the amplitude of the noise, but most importantly how large is the change of the spectra under a variation of the parameters compared to the noise (i.e. the Fisher Matrix), and how different the effect of each parameter is on the spectra, so that degeneracies can be broken. In the case of  \TE, the noise is indeed relatively high, but this spectrum is very efficient at breaking degeneracies. This can be inferred by comparing the constraints from  \TE\  in the diagonal case in Fig. \ref{diaglensedCVL} with the ones in the reference case in Fig.                              \ref{lensedCVL}. In the diagonal case, where the degeneracy breaking power of each spectrum is removed,  \TE\  becomes much worse at constraining $\ns$, $\lnA$ and $\tau$ than  \TT\  or  \EE\ alone, and it becomes comparable to  \TT\  on $\Omega_ch^2$. However, it remains very good at constraining $\theta$ and $\Omega_bh^2$ because intrinsically changing these parameters has a large impact on the cross-correlation spectrum.

\subsection{Evolution of the constraints with \lmin}
\label{sectionlmin}
\begin{figure}[t]
\centering
 \includegraphics[angle=0,width=0.5\textwidth]{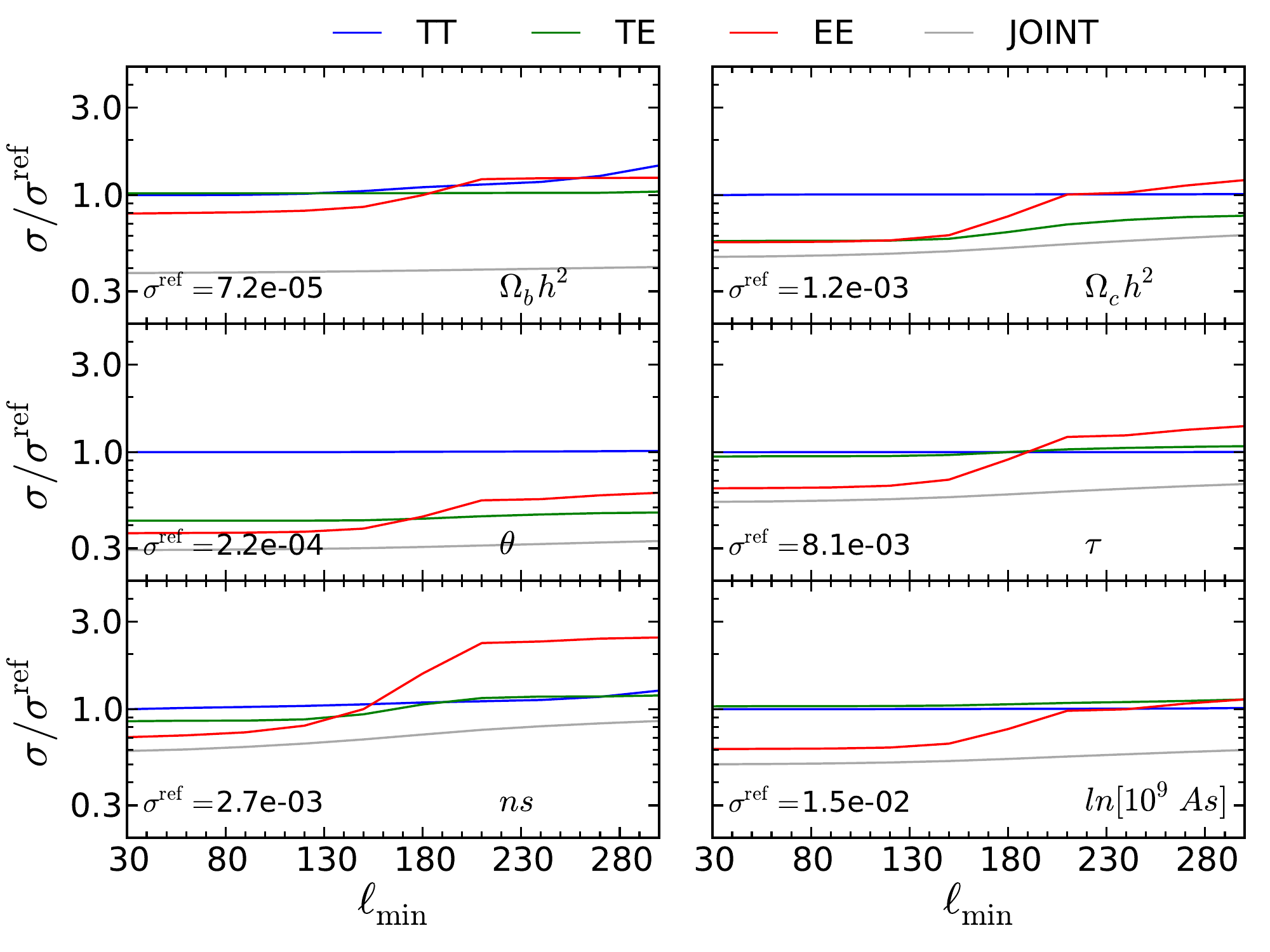}
\caption{Standard deviations on $\Lambda CDM$ parameters as function of \lmin, normalized to the standard deviation $\sigma^{\rm ref}$ obtained from \TT\ with $\lmin=30$. We consider a \CVL\ experiment with $\lmax=2500$ and a prior on $\tau$. We consider here lensed CMB power spectra.}
\label{lensedCVLlmin}
\end{figure}

\begin{figure}[t]
\centering
 \includegraphics[angle=0,width=0.5\textwidth]{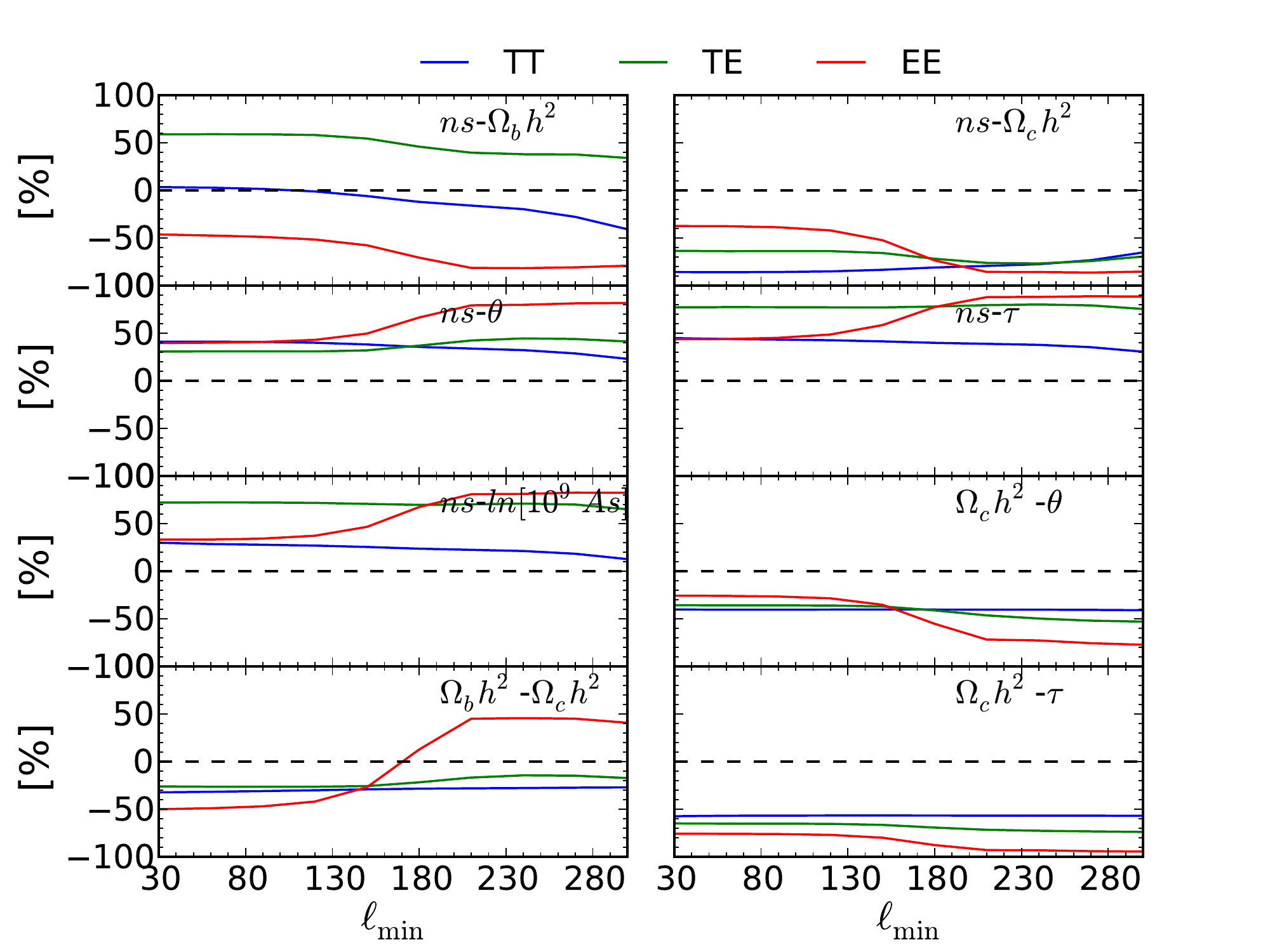}\caption{Evolution of the correlations between selected parameters as a functions of \lmin. We consider a \CVL\  experiment with $\lmax=2500$ and a prior on $\tau$. We consider here lensed CMB power spectra.}
\label{corrlminCVL}
\end{figure}

Fig. \ref{lensedCVLlmin} shows the evolution of the constraints with  $\lmin$, varied between $30<\lmin<300$ at a fixed $\lmax=2500$. In the \CVL\  case  considered here, we naively expect that cutting the noisier low-$\ell$ part of the spectrum should only marginally affect the results. Indeed, we find that the \TT\  constraints are marginally affected even when the most dramatic cut ($\lmin=300$) is applied. Only the constraint on $\Omega_b h^2$ is worsened by $\sim 50\%$ for $\lmin \gtrsim 200$, due to the fact that with such a cut,  the first peak is not observed anymore, reducing the information on the baryon load coming from the difference in height between the first and the second peak \cite{hu95}.

On the other hand, the constraints from the \EE\ polarization are more drastically affected by cutting the low-$\ell$ part. In particular, there is a strong worsening  of the constraints of all parameters when $\lmin\gtrsim 130$. This is due to the fact that the inclusion of the multipoles  $\ell \lesssim 200$  alleviates degeneracies, in particular between the scalar spectral index $\ns$ and other parameters. This is clear from Fig. \ref{corrlminCVL}, where we plot the correlation coefficients between couples of  parameters (mainly $\ns$ versus the others) as a function of $\lmin$. For $\lmin$ larger than $\lmin \gtrsim 130$, $\ns$ becomes almost $ \sim \pm 100\%$ correlated with the other parameters for  \EE.
Furthermore,  we verified that if we fix $\ns$ (thus marginalizing over only 5 parameters), this sharp worsening of the constraints disappears, while if we fix any of the other parameters, the step is still present, although  when fixing $\Omega_ch^2$  it appears only in the constraints for $\Omega_bh^2$ and $\ns$. This suggests that cutting at $\lmin\gtrsim 130$ largely impacts the constraints from  \EE\ due to the increasing  degeneracies between $\ns$ and other parameters, in particular with $\Omega_ch^2$ and $\Omega_bh^2$.

The reason why this $\ell-$range helps breaking such degeneracies can be inferred from  Fig. \ref{derivativesEE}, that shows the derivatives of the \EE\ power spectrum relative to the different $\Lambda$CDM parameters. The derivatives with respect to $\Omega_bh^2$ and $\Omega_ch^2$ change behaviour at $\ell\sim200$, thus helping to break degeneracies. 
The physical reason of this is the following. 
\smallskip

\noindent\textbf{Breaking the $\Omega_ch^2-\ns$ degeneracy.} Scales below $\ell \lesssim 200$ entered the causal horizon well after matter-radiation equality ($z_{eq}\sim 3400$). 
These scales are thus not affected by the decay of the metric perturbations during radiation domination, that act as a driving force on the acoustic oscillations, boosting their amplitude \cite{hu95}. Increasing $\Omega_mh^2$ by increasing $\Omega_ch^2$ anticipates the redshift of equality, and thus decreases the boosting effect due to the driving, shifting it to smaller scales. The overall result is thus a decrease in the amplitude of the oscillations at $\ell\gtrsim 200$, and an almost negligible change at larger scales. This is then reflected in a negative derivative relative to $\Omega_ch^2$ at $\ell\gtrsim 200$, and in a very small one, (by almost two orders of magnitude) at larger scales. This difference in behaviour of the spectrum under a change in $\Omega_ch^2$ at large and small scales significantly helps breaking degeneracies with $\ns$, which smoothly changes the spectrum at these scales.

It is also interesting to notice that such a difference is not so evident in the derivative of the   \TT\  spectrum relative to  $\Omega_ch^2$, as shown in Fig. \ref{derivativesEE}. This is due to the fact that an increase in the matter density also determines a decrease of the early and late Integrated Sachs-Wolfe effects (ISW) at large scales in temperature (these effects are absent in polarization), so that the derivatives are negative and large even at these scales. Therefore,  degeneracies are not particularly broken by observing this $\ell-$range in  \TT.
\smallskip

\noindent\textbf{Breaking the $\Omega_bh^2-\ns$ degeneracy.}
As far as the baryon density is concerned, changing $\Omega_bh^2$ affects  \EE\ through the change in $\Omega_mh^2$, similarly to $\Omega_ch^2$. However, in this case, there are also other, more relevant effects. The first is that increasing $\Omega_bh^2$  decreases the sound speed $c_s=1/\sqrt{(3(1+R))},\, R=\frac{3\Omega_b a}{4\Omega_r} $.
As the amplitude of the temperature dipole, sourcing the temperature quadrupole and thus polarization, is proportional to $c_s$, an increase in $\Omega_bh^2$ causes a decrease in the amplitude of the  \EE\ spectrum, proportional to $c_s^2$ \cite{hu95}. This effect acts in principle on all  scales smaller than the causal horizon at the epoch of decoupling, $\ell\lesssim 100$, and would result in a negative derivative relative to $\Omega_bh^2$ at all these scales. However, Fig. \ref{derivativesEE} shows that the derivative at $\ell\lesssim 200$ is positive. This is due to the fact  that increasing $\Omega_bh^2$ increases the effective mass of the photo-baryonic fluid, enhancing the gravitational potential wells. This  increases  the amplitude of the  spectra (both temperature and polarization) by a factor $(1+3R)^2$ \cite{huthesis,hu95}, and it is the dominant effect at large scales, where the metric perturbations have not decayed during radiation domination. This explains why the derivative relative to $\Omega_bh^2$ is positive at these scales\footnote{Let us mention here, for completeness, that the derivatives of  all the spectra relative to $\Omega_bh^2$ are positive also at $\ell\gtrsim 1000$. This is due to the fact that increasing the baryon density decreases the Silk damping \cite{silk68, kaiser83}, as the Silk damping scale $k_D$ is proportional to $k_D\propto \sqrt{(\Omega_bh^2)}$. We also verified that the changes in the recombination history and on the visibility function due to a change in $\Omega_bh^2$ have minor effects on the derivatives.
Finally, the derivatives relative to $\Omega_ch^2$ also change behavior at $\ell\gtrsim 1000$. This is due to the effect of lensing: increasing the amount of matter increases the amount of lensing, thus increasing the power and the smoothing of the peaks/troughs at high-$\ell$ \cite{zaldarriaga98}.}. Notice that the effect of baryons on  \TT\  has additional features compared to  \EE:  the change in the effective mass also produces a shift in the equilibrium point of the oscillations, resulting in the enhancement of the odd peaks (compressions of overdensities) compared to even ones (rarefactions of overdensities), at scales where the metric perturbation have not completely decayed during radiation domination. This effect dominates over the decrease in the sound speed, that affects  \TT\  reducing the amplitude of the Doppler contribution to the oscillations.

\section{Planck}
\label{sec:planck}
 \begin{figure*}[ht!]
\centering
 \includegraphics[angle=0,width=0.7\textwidth]
{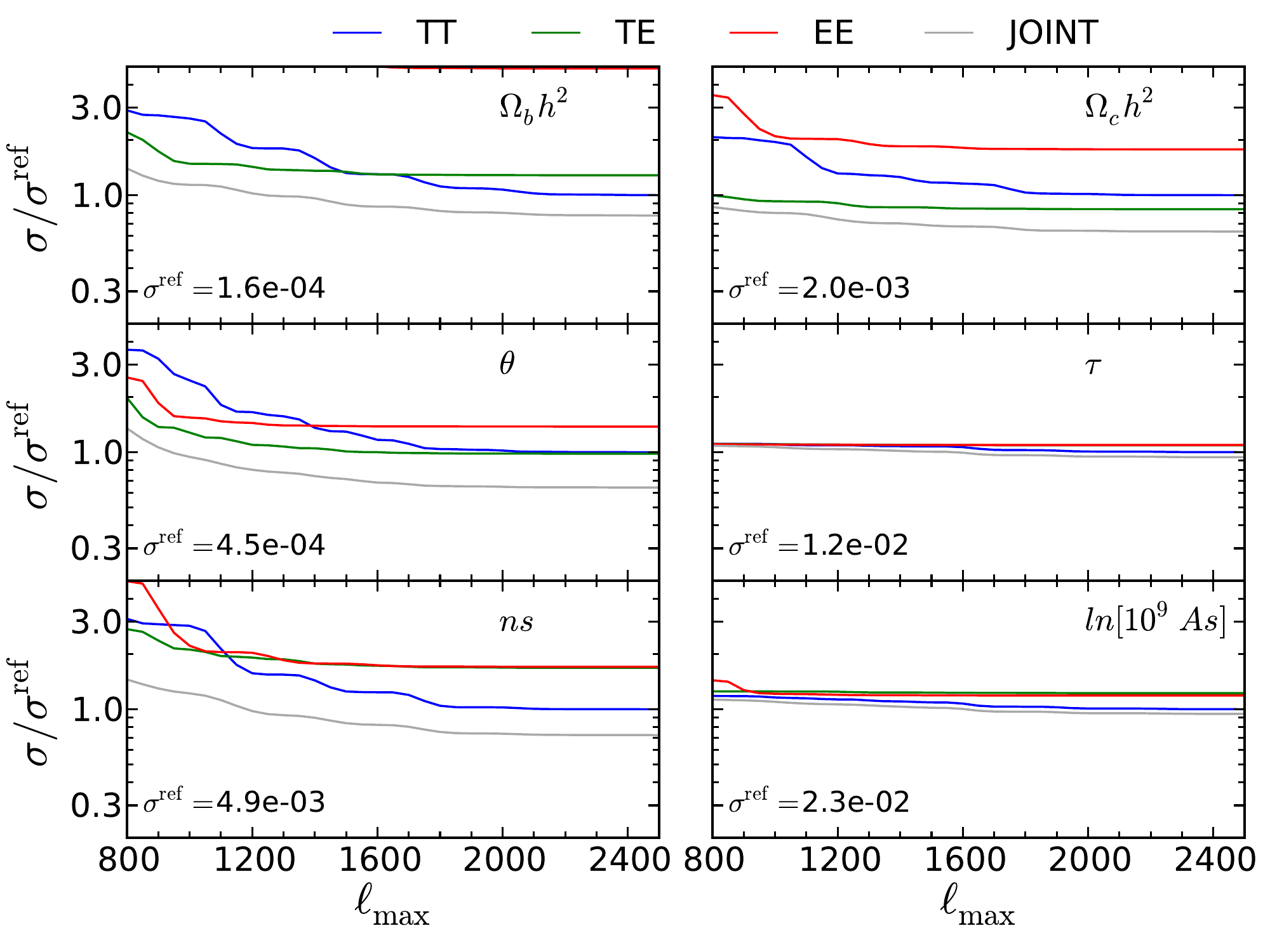}
%resultsPlanck2chanFull_fsky_0.5_tauprior_lensed_6par_varlmax/errors_Planck2chanFull_tauprior_lmin30_lmax2500_noisefac1_varlmax.pdf}
\caption{Standard deviations on $\Lambda CDM$ parameters as function of \lmax, normalized to the standard deviation $\sigma^{\rm ref}$ obtained from \TT\ with $\lmax=2500$. We consider a Planck-like full mission experiment with two channels at $143$ GHz and $217$ GHz with $\lmin=30$ and a prior on $\tau$. We consider here lensed CMB power spectra.}
\label{plancklmax}
\end{figure*}

\begin{figure}[h]
\centering
%resultsPlanck2chanFull_fsky_0.5_tauprior_lensed_6par_varlmax/errors_Planck2chanFull_tauprior_lmin30_lmax2500_noisefac1_varlmax.pdf}
 \includegraphics[angle=0,width=0.5\textwidth]
{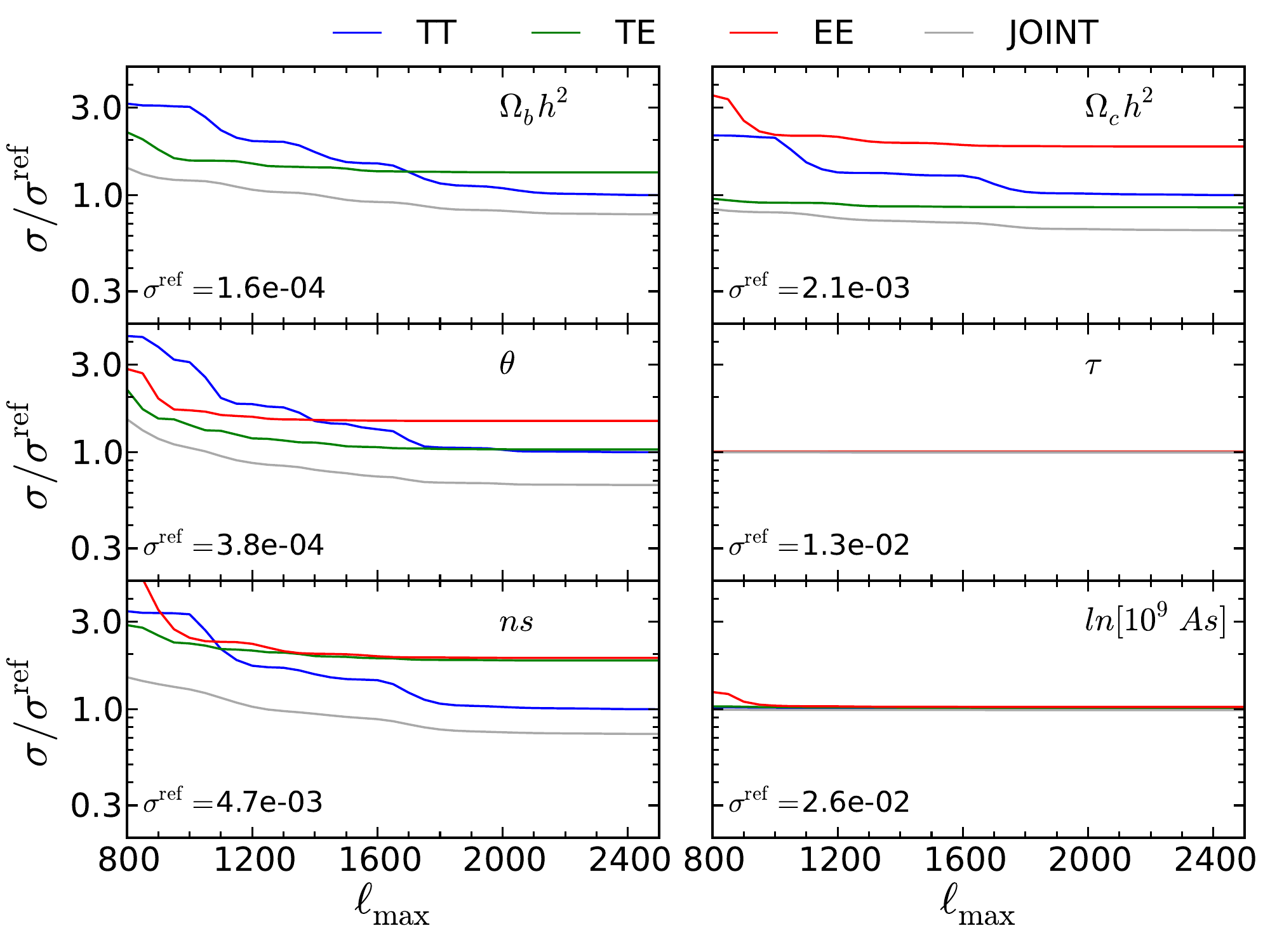}
\caption{Same as Fig. \ref{plancklmax}, but for unlensed spectra.}
\label{scalarplancklmax}
\end{figure}

\begin{figure}[t]
\centering
 \includegraphics[angle=0,width=0.5\textwidth]
{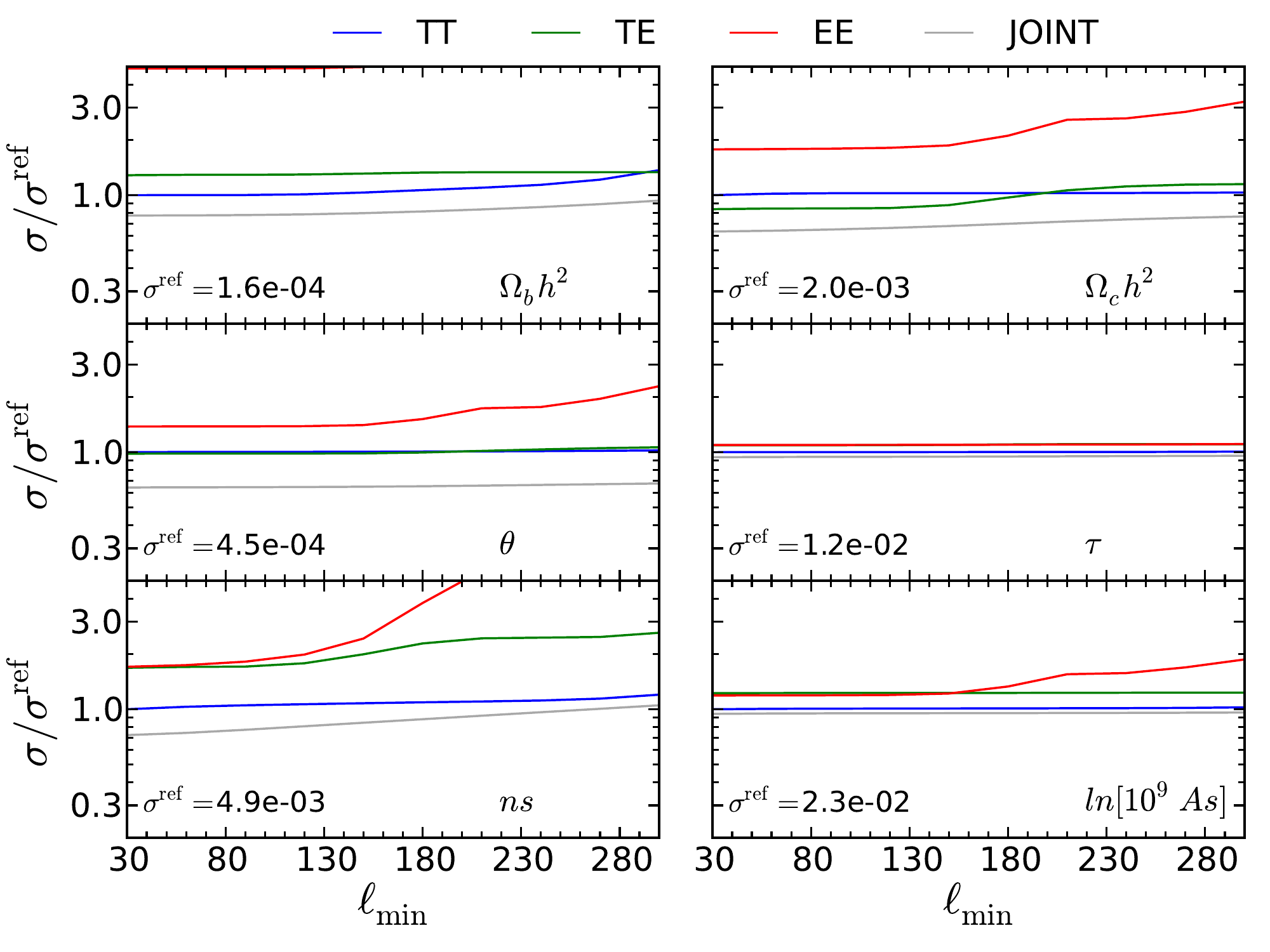}
 \caption{Standard deviations on $\Lambda CDM$ parameters as function of \lmin, normalized to the standard deviation $\sigma^{\rm ref}$ obtained from \TT\ with $\lmin=30$. We consider a Planck-like full mission experiment $\lmax=2500$ and a prior on $\tau$. We consider here lensed CMB power spectra.}
\label{Plancklmin}
\end{figure}

In this section, we repeat the analysis  performed in the previous Sections for the case of a Planck-like experiment, assuming full mission data ($30$ months of data).
We combine the $143$ and $217$ GHz channels, and assume specifications as in Table \ref{tab:exp}, following \cite{blubook}. Clearly, the forecasts in this Section are overly simplified, as we e.g. ignore  the presence of foregrounds, use only two frequency channels, ignore $\ell$ by $\ell$ correlations due to the galactic/point source masking of the maps. However, we verified that for a Planck-like nominal mission experiment ($14$ months of data), the error bars obtained from our Fisher Matrix forecast for  \TT\   alone are similar to the ones obtained using the published Planck data from the first data-release \cite{planckcosmo}, within a factor $1.3-1.5$. We therefore expect the forecasts to be in the right ballpark.
%\subsection{Evolution of the constraints with lmax}

Figure \ref{plancklmax} shows the evolution of the constraints on cosmological parameters as a function of \lmax, while Fig. \ref{scalarplancklmax} shows the same evolution but for unlensed spectra. 

The most interesting result in Figure \ref{plancklmax} is that we forecast that \TE\  should provide a constraint on the dark matter density $\Omega_ch^2$ stronger than the one obtained from \TT\  alone, namely $\sigma_{TE}(\Omega_ch^2)=1.7\times 10^{-3}$ [1.4\%] versus $\sigma_{TT}(\Omega_ch^2)=2.0\times 10^{-3}$ [1.7\%], while providing constraints comparable to \TT\  for $\Omega_bh^2$, $n_s$ and $\theta$, as also shown in Table \ref{tab:planck}. We underline that this is the first time that the power of  \TE\  alone  is forecasted in a paper.
 
\begin{table*}[thb]%\footnotesize
\begin{center}
\begin{tabular}{rcccccccccccc}
\hline\hline
\noalign{\vskip 3pt}
Data & $\Omega_bh^2$ & & $\Omega_ch^2$ & & $\theta$ & & $\tau$ & & $\ns$ & & $\lnA$\\ 
\noalign{\vskip 3pt}
\hline
\noalign{\vskip 3pt}
TT  & $1.7 \times 10^{-4}$ & [0.8\%] & $2.0\times 10^{-3}$ & [1.7\%] & $4.5\times 10^{-4}$ & [0.04\%] & $1.2\times 10^{-2}$ & [13\%] & $4.9\times 10^{-3}$ & [0.5\%] & $2.3\times 10^{-2}$ & [0.7\%] \\
EE  & $8.0 \times 10^{-4}$ & (0.2)   & $3.6\times 10^{-3}$ & (0.56)  & $6.1\times 10^{-4}$ & (0.72)   & $1.3\times 10^{-2}$ & (0.91) & $8.3\times 10^{-3}$ & (0.59)  & $2.7\times 10^{-2}$ & (0.84)  \\
TE  & $2.1 \times 10^{-4}$ & (0.78)  & $1.7\times 10^{-3}$ & (1.2)   & $4.4\times 10^{-4}$ & (1.0)    & $1.3\times 10^{-2}$ & (0.92) & $8.2\times 10^{-3}$ & (0.59)  & $2.8\times 10^{-2}$ & (0.82)  \\
JOINT & $1.3 \times 10^{-4}$ & (1.3)   & $1.3\times 10^{-3}$ & (1.6)   & $2.9\times 10^{-4}$ & (1.6)    & $1.1\times 10^{-2}$ & (1.1)  & $3.5\times 10^{-3}$ & (1.4)   & $2.1\times 10^{-2}$ & (1.1)   \\
\noalign{\vskip 3pt}
\hline
\end{tabular}

\caption{Standard deviations on the $\Lambda$CDM model for a Planck-like full mission experiment from  \TT\,  \EE\,  \TE\ taken separately or from the combination of all the power spectra. These constraints are calculated assuming $\lmin=30$, $\lmax=2500$, and a prior on $\tau$, $\sigma(\tau)=0.013$. In square brackets, on the first line we translate the standard deviation in relative error. In parenthesis, on the next lines we show the improvement factor compared to the  \TT\  case.}

\label{tab:planck}
\end{center}
\end{table*}

This opens the  interesting possibility of cross-checking the results obtained from  \TT\  alone, also considering  that  \TE\ is expected to have a lower level of foreground contamination at small  scales than  \TT. Furthermore, the constraints from  \TE\ do not seem to improve extending $\lmax \gtrsim 1500$, as shown in Fig. \ref{plancklmax}, or by cutting  at $\lmin\lesssim 150$, as shown in Fig. \ref{Plancklmin}. This provides the possibility of obtaining strong constraints even when eliminating the most problematic multipole ranges, where polarized contamination from the galaxy (at large scales) or from extragalactic foregrounds (at small scales) are expected to be most relevant.

Furthermore, comparing Fig. \ref{plancklmax} (from lensed spectra) with Fig. \ref{scalarplancklmax} (from unlensed spectra), we  notice that the degeneracy-breaking effect of lensing on $\lnA-\tau$  is effective only for  \TT\  alone, as  \TE\ and  \EE\ apparently do not have a high enough signal-to-noise ratio to sufficiently observe such an effect at small scales, as also shown in Fig. \ref{spectra}.  Moreover, the constraints on the other parameters are only  marginally affected by the additional information provided by lensing, as also shown in Table \ref{lensedscalarratio}.
Figure \ref{Plancklmin} shows the evolution of the constraints with $\lmin$. As previously noted for the \CVL\  case in Section \ref{sectionlmin}, the constraints from \EE\ rapidly worsen when $\lmin \gtrsim 130$, due to the fact that multipoles between $ 130\lesssim\ell\lesssim200$ help breaking degeneracies between $\ns$ and other parameters. We notice in particular that $\ns$ worsens by a factor $\sim 3$ if we cut $\lmin=200$. This indicates that for  \EE, the proper handling of the galactic dust contamination at large scales is crucial to have accurate results, as the observation in this $\ell-$range has  such large impact on the constraints. 

\section{The effect of  a prior on $\tau$}
\label{sec:tau}
\begin{figure*}[h!]
\centering
\includegraphics[angle=0,width=0.45\textwidth]{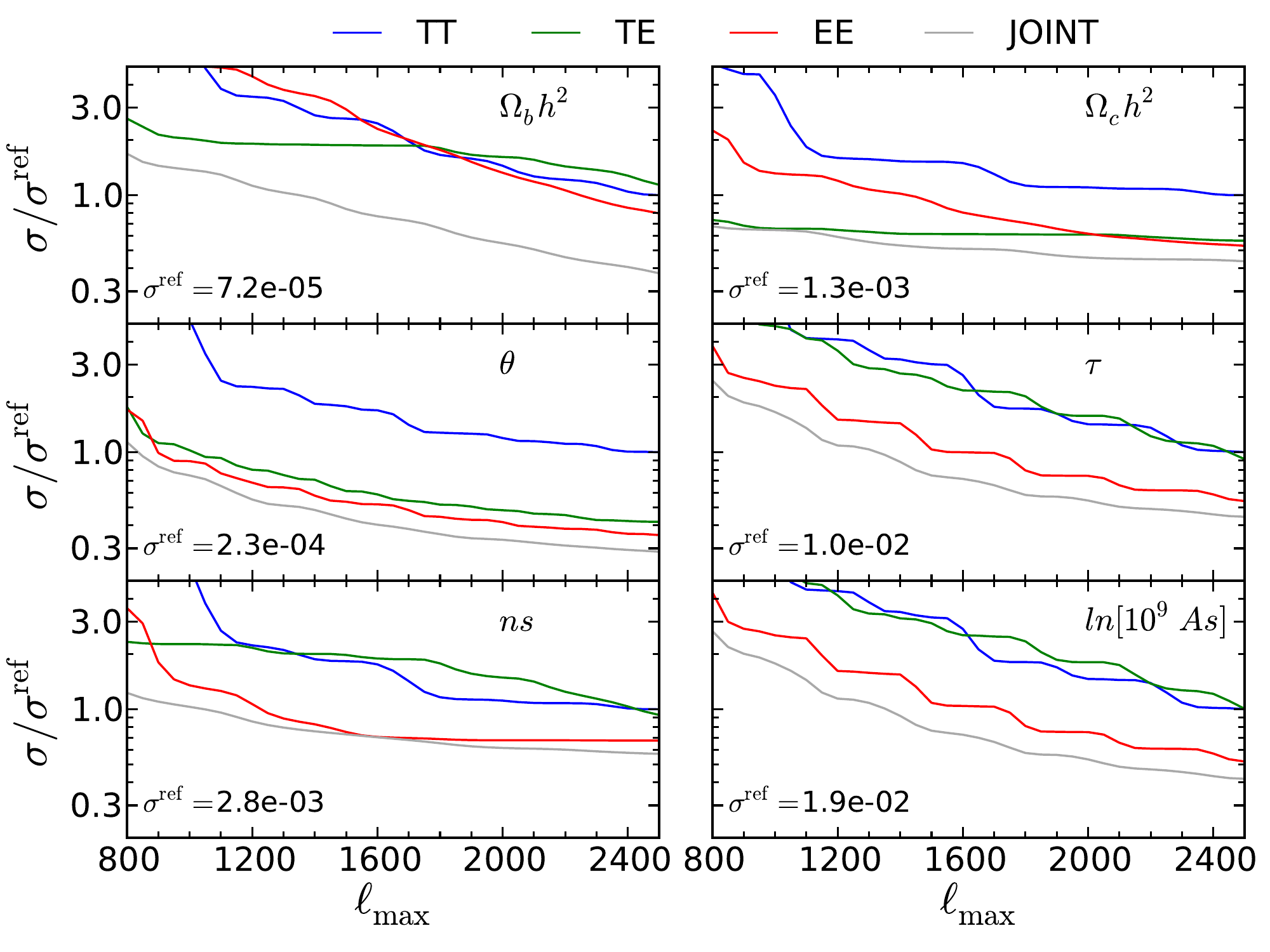}
\includegraphics[angle=0,width=0.45\textwidth]{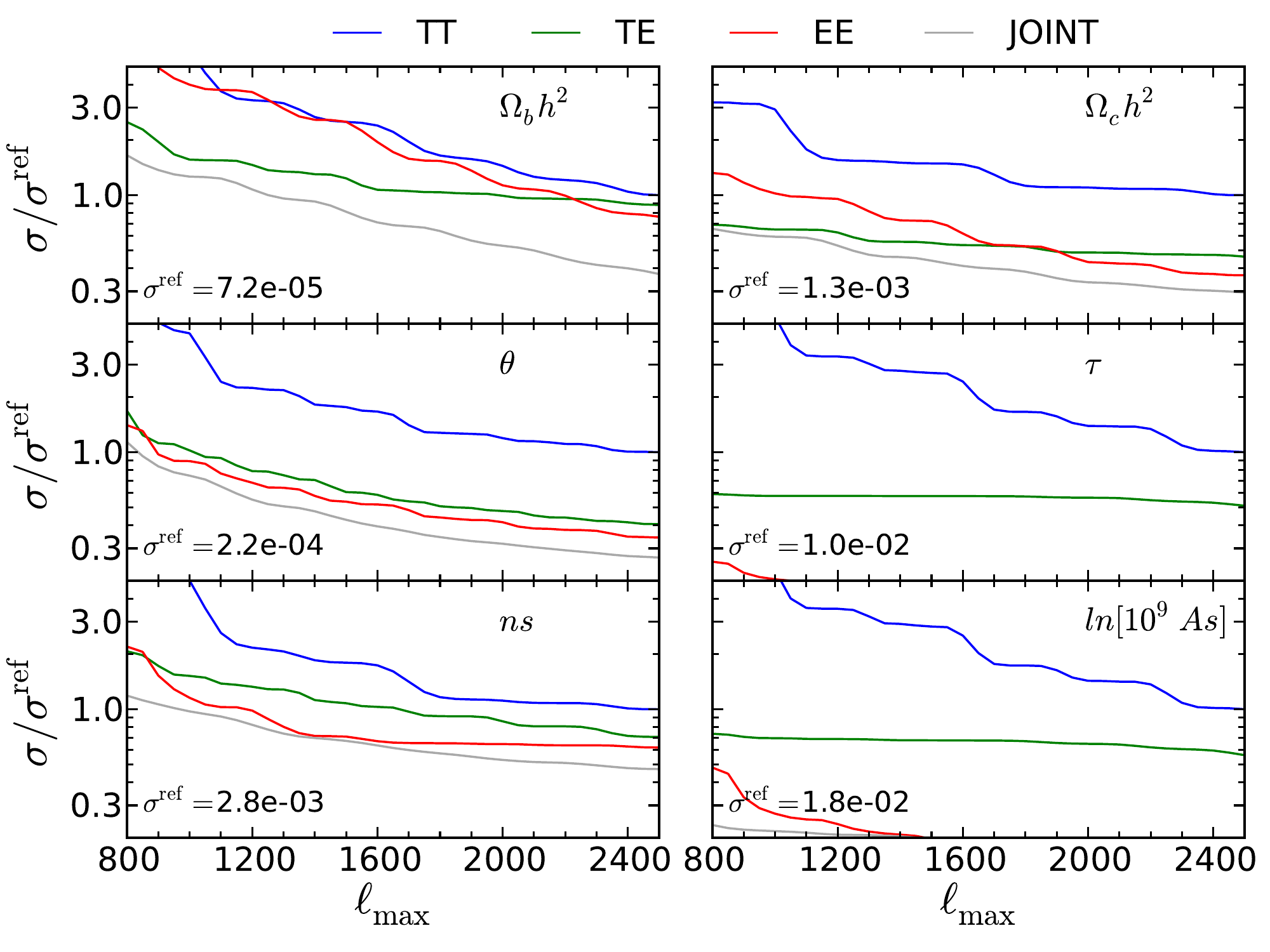}
\caption{Standard deviations on $\Lambda CDM$ parameters as function of \lmax\ {\it without} any prior on $\tau$, normalized to the standard deviation $\sigma^{\rm ref}$ obtained from \TT\ with $\lmax=2500$.  We consider a \CVL\ experiment with $\lmin=30$ (left) and with $\lmin=2$ (right). We consider here lensed CMB power spectra. }
\label{lensedCVLlmaxnNOTAU}
\end{figure*}

In the previous sections we forecasted constraints always assuming a prior on  $\tau$. In this Section we want to analyze whether relaxing this assumption impacts our  conclusions. 

Fig. \ref{lensedCVLlmaxnNOTAU} (left plot) shows the evolution of the constraints as a function of $\lmax$ without including any prior on $\tau$ and fixing $\lmin=30$. Comparing to Figure \ref{lensedCVL}, we  notice that, as expected, the uncertainties on parameters increase by a factor of $1.2-3$, due to the now much wider degeneracy between $\tau$ and other parameters (in particular $\lnA$). However, our previous conclusions do not qualitatively change, i.e. that the \EE\ power spectrum is  the best at constraining parameters when the full $\ell-$range is considered. 

Furthermore,  we notice that thanks to the degeneracy-breaking effect of lensing, including $\lmax=2500$ the constraint on  $\tau$ from \TT\  alone is $\sigma_{TT}({\tau})=0.010$ [11\%], from \TE\ is $\sigma_{TE}({\tau})=0.010$ [11\%] and from \EE\ is $\sigma_{EE}({\tau})={0.006}$ [6\%]. This can be compared to the constraints obtainable from  polarization when the low-$\ell$ part of the spectrum is included in the analysis, i.e. when the reionization bump is included. This is shown in Fig. \ref{lensedCVLlmaxnNOTAU} (right plot), where we calculate constraints assuming $\lmin=2$. In this case we obtain from \TE\  $\sigma_{TE}({\tau})=0.005$ [5\%] and from \EE\  $\sigma_{EE}({\tau})={0.002}$ [2\%]. This indicates that observing the reionization bump can place constraints on $\tau$ that are stronger by about a factor $3$ compared to the ones obtained using the degeneracy-breaking effect of lensing alone\footnote{Notice that when we do not use a $\tau$ prior, the degeneracy with $\lnA$ becomes very large, and the likelihood distribution of the parameters might become non-gaussian, rendering the estimates from the Fisher Matrices overly optimistic. }. Furthermore, both in the cases when $\lmin=2$ or $\lmin=30$,  \EE\ is best at constraining $\tau$, either from the large scale reionization bump (as already noticed in \cite{Kaplinghat03}) or from the degeneracy-breaking effect of lensing.

\begin{figure*}[t]
\centering
\includegraphics[angle=0,width=0.45\textwidth]{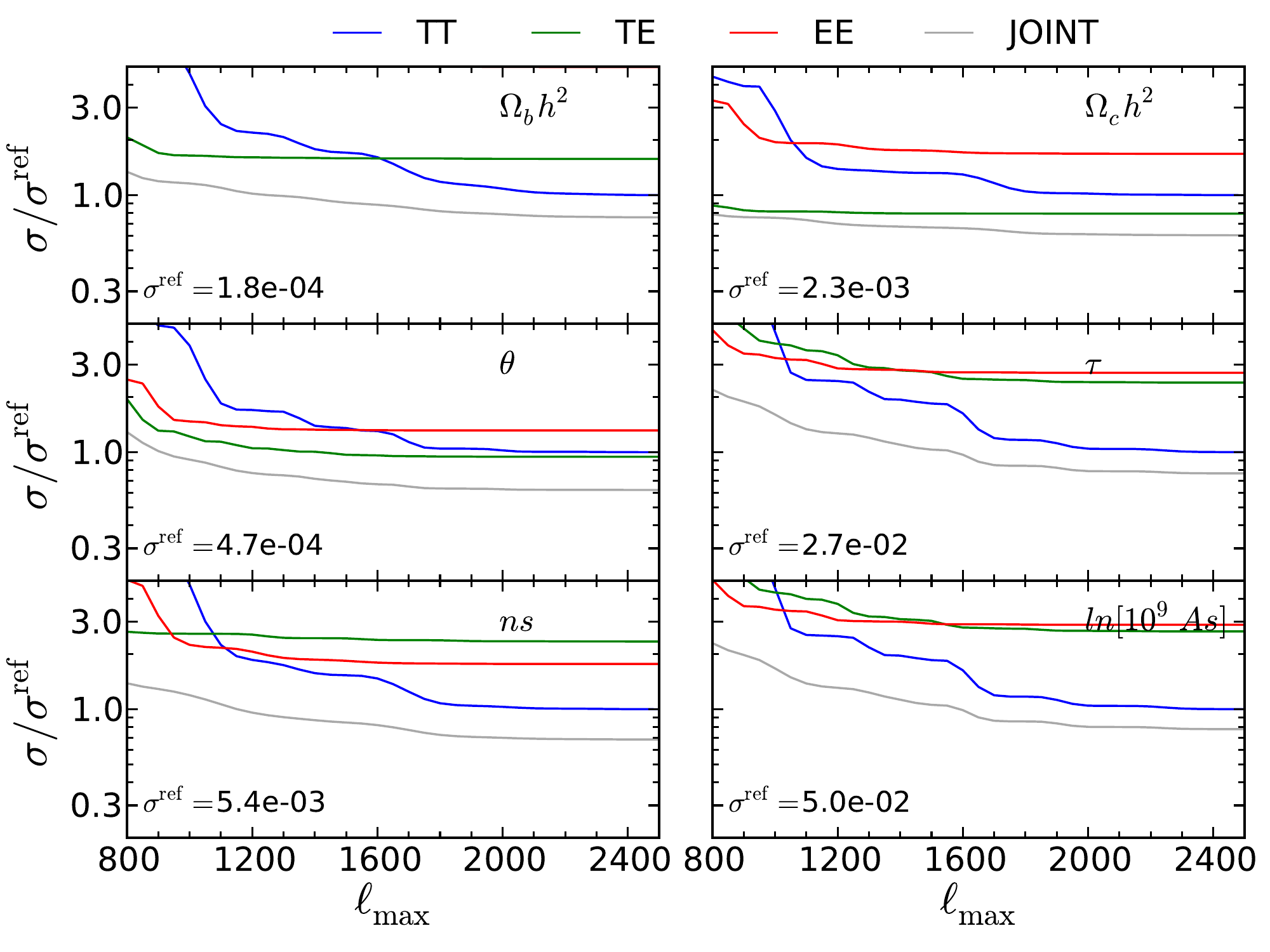}
\includegraphics[angle=0,width=0.45\textwidth]{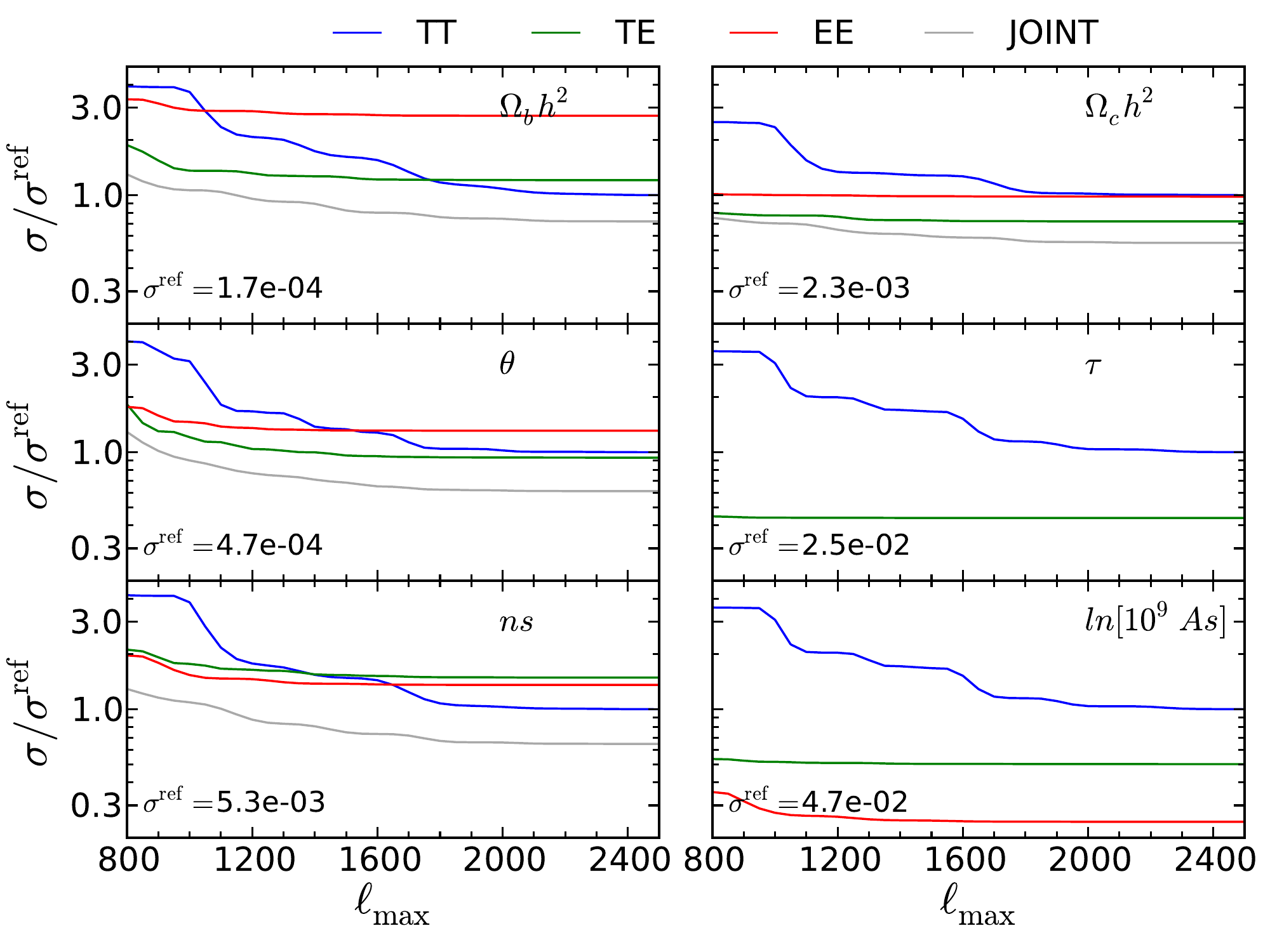}
\caption{Standard deviations on $\Lambda CDM$ parameters as function of \lmax, normalized to the standard deviation $\sigma^{\rm ref}$ obtained from \TT\ with $\lmax=2500$. We consider a Planck-like experiment with $\lmin=30$ (left) and $\lmin=2$ (right), {\it without} any prior on $\tau$. We consider here lensed CMB power spectra.}
\label{lensedplancklmaxnNOTAU}
\end{figure*}

We perform this same analysis also for a Planck-like full mission experiment, with specifications already detailed in Section \ref{sec:planck}.
 Fig. \ref{lensedplancklmaxnNOTAU} (left plot) shows the constraints on parameters without any prior on $\tau$ and with $\lmin=30$. As already noticed in Section \ref{sec:planck}, only temperature has enough signal-to-noise ratio to take advantage of the degeneracy breaking power of lensing, achieving a constraint on $\tau$ of $\sigma_{TT}(\tau)=0.027$ [29\%]. This can be compared to the constraint obtained from  \EE\ alone,  $\sigma_{EE}(\tau)=0.072$ [78\%] and from  \TE\ alone, $\sigma_{TE}(\tau)=0.064$ [69\%].
 On the other hand, when including the low-$\ell$ part ($\lmin=2$), as shown in Fig. \ref{lensedplancklmaxnNOTAU} (right plot), we again see that the  \EE\ power spectrum is the best at constraining $\tau$. In particular, we obtain from  \EE\ $\sigma_{EE}(\tau)=0.0046$ [5\%], that compared to the results from  \TE, $\sigma_{TE}(\tau)=0.011$ [12\%] are a factor $\sim 2$ stronger. These forecasts however  do not take into account the difficulties in dealing with the contamination of the galactic emission or systematics, that might considerably worsen  the expected constraints from the reionization bump.
 
\section{$\Lambda$CDM extensions}
\label{sec:exten}
In this Section, we present the constraints on extensions of the $\Lambda$CDM model. In particular, we explore cases where we constrain one additional parameter at a time: the sum of neutrino mass $\sum m_\nu$, the number of relativistic species $\neff$, the Helium abundance $Y_p$ or running of the scalar spectral index $n_{run}$.
Current constraints from Planck limit the sum of the neutrino mass to be smaller than $\sim 1eV$, so that neutrinos are expected to become non-relativistic only after the baryon-photon decoupling. Thus, the sum of the neutrino masses affects only the late time evolution of the spectra, changing the early and late ISW effects through a change in the expansion history in the  \TT\  power spectrum at large scales, and by changing the amount of lensing at small scales (a larger neutrino mass smooths the matter power spectrum at scales smaller than the free-streaming scales, decreasing the effect of lensing) \cite{hou14}. 

The number of relativistic species on the other hand affects the spectra in a number of different ways \cite{WMAP7,hou11,bashinsky04}, and we describe only the main ones. An increase of $\neff$ determines an increase of the radiation density, so that matter-radiation equality happens at later times. This increases the amplitude of acoustic peaks in all spectra through an increase of the radiation driving, as explained in Section \ref{sectionlmin}, and at large scales in  \TT\  through an increase of early ISW. This effect is strongly degenerate with $\Omega_ch^2$. Furthermore, it increases the angular Silk damping scale $\theta_D\propto r_D/D_*$ relative to the angular size of sound horizon $\theta$, with $r_D$ the comoving Silk damping scale and $D_*$ the comoving distance to decoupling. In fact, having $\theta$ fixed, $\theta_D$ is proportional to the square of the Hubble parameter $H^{0.5}$ \cite{hou11}, so that an increase of the radiation density increases $H\sim \sqrt{((\Omega_ch^2+\Omega_b^2)a^{-3}+\Omega_rh^2a^{-4})}$ and therefore $\theta_D$, i.e. it increases the effect of Silk damping shifting it to larger scales.

A change in the primordial Helium abundance $Y_p$ changes the epoch of recombination: a larger amount of helium relative to hydrogen leaves a smaller amount of free electrons after helium recombination, so that hydrogen recombination happens earlier, and the width of the last scattering surface is larger, enhancing Silk damping \cite{ichikawa08}.

Finally, changing $n_{\rm run}$ changes the relative amount of power at large/small scales, as we define the power spectrum as  $$P_R(k) = A_s \left(\frac{k}{k_0}\right)^{n_s-1+(1/2)(dn_{\rm s}/d\ln{k})({\rm ln}{(k/k_0)})},$$with $k_0=0.05 \, {\rm Mpc}^{-1}$ the pivot scale and $dn_{\rm s}/d{\rm ln}{k}\equiv n_{\rm run}$ the running.

\begin{figure*}[tbh]
\centering
\includegraphics[angle=0,width=0.35\textwidth]{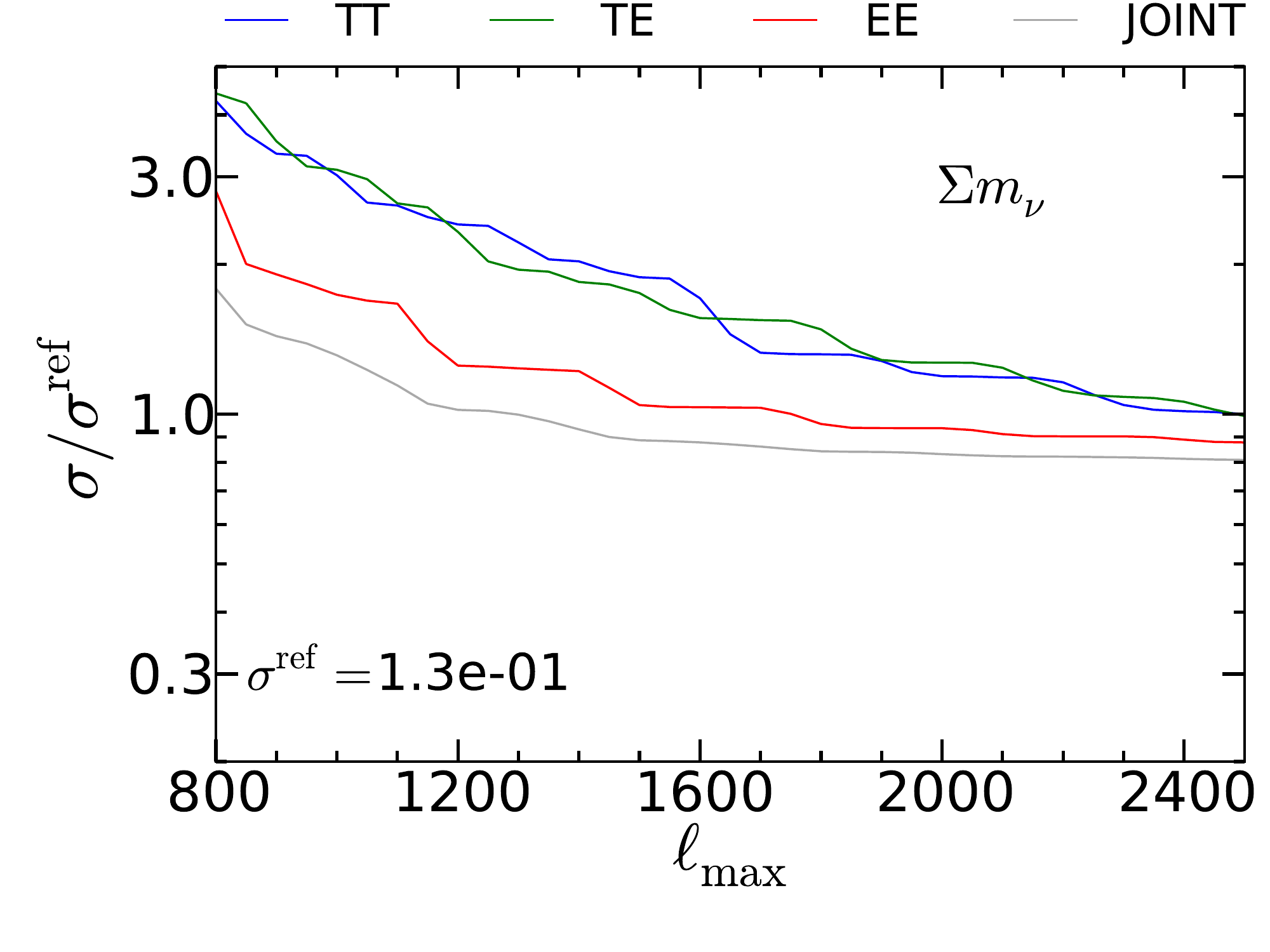}
\includegraphics[angle=0,width=0.35\textwidth]{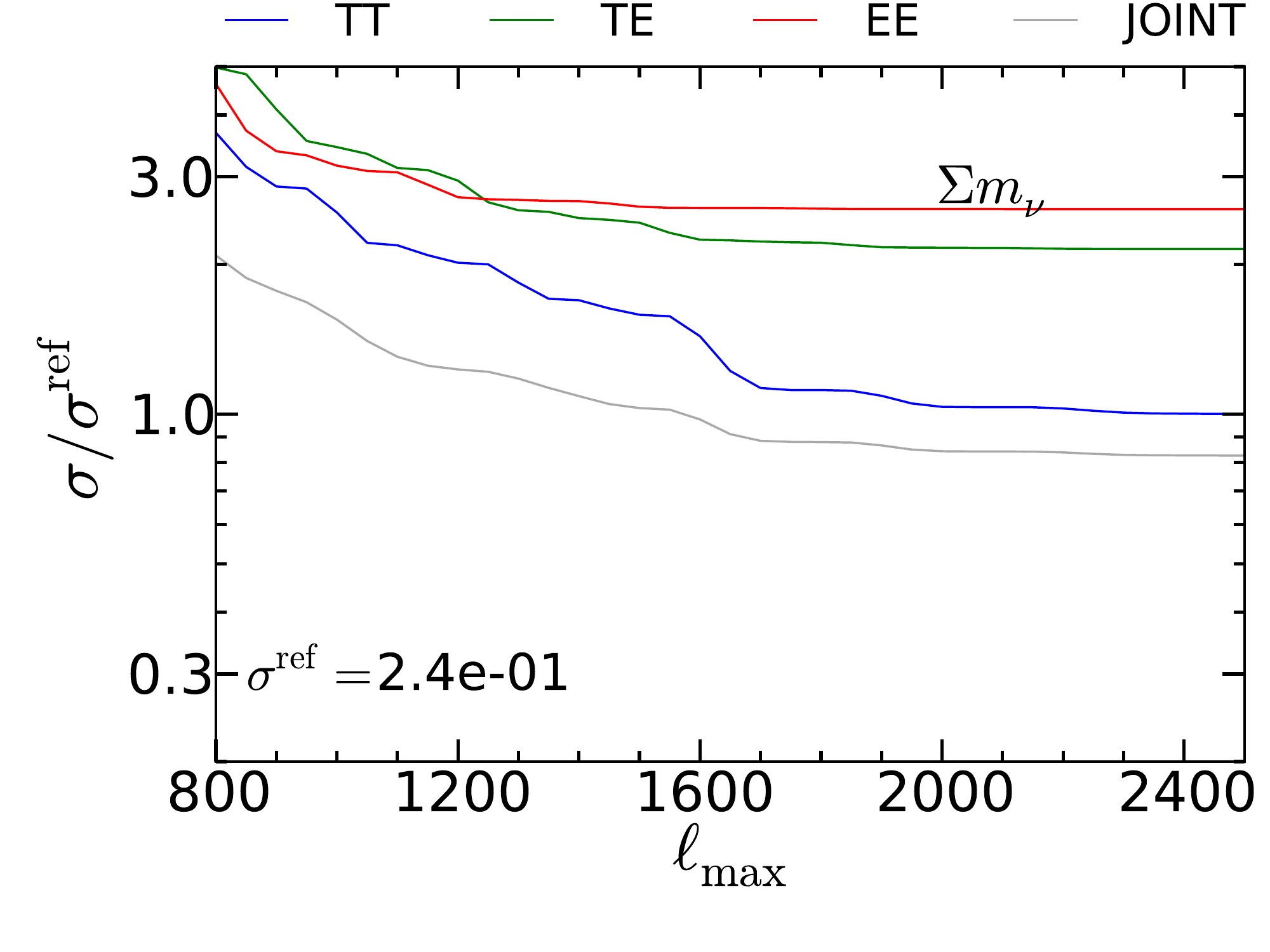}
\caption{Standard deviations on $\Sigma m_\nu$ as function of $\lmax$, normalized to the standard deviation $\sigma^{\rm ref}$ obtained from \TT\ with $\lmax=2500$. We consider a \CVL\ experiment (left) and a Planck-like full mission experiment (right) with $\lmin=30$ and a prior on $\tau$. }
\label{mnu}
\end{figure*}

\begin{figure*}[tbh]
\centering
\includegraphics[angle=0,width=0.35\textwidth]{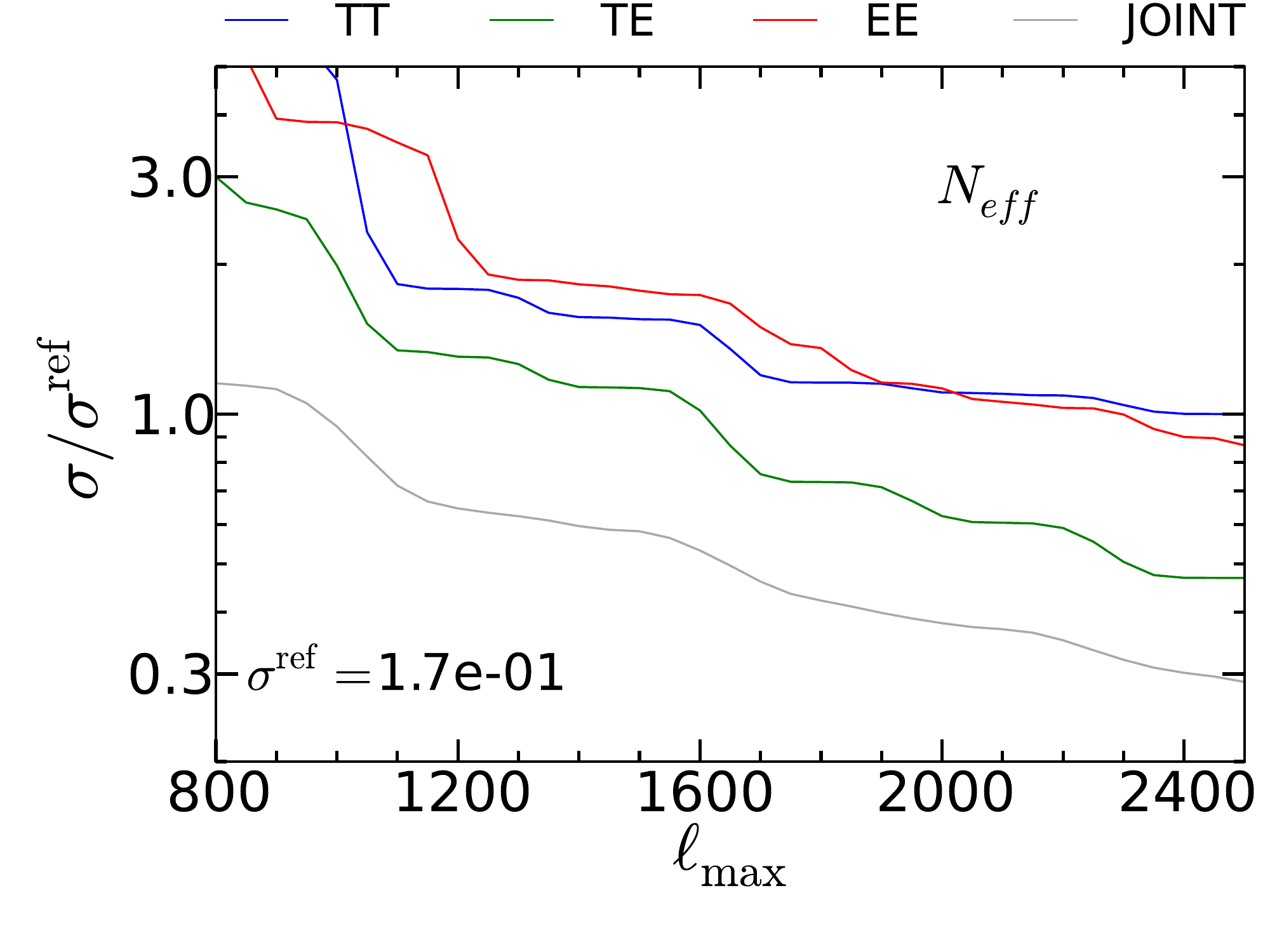}
\includegraphics[angle=0,width=0.35\textwidth]{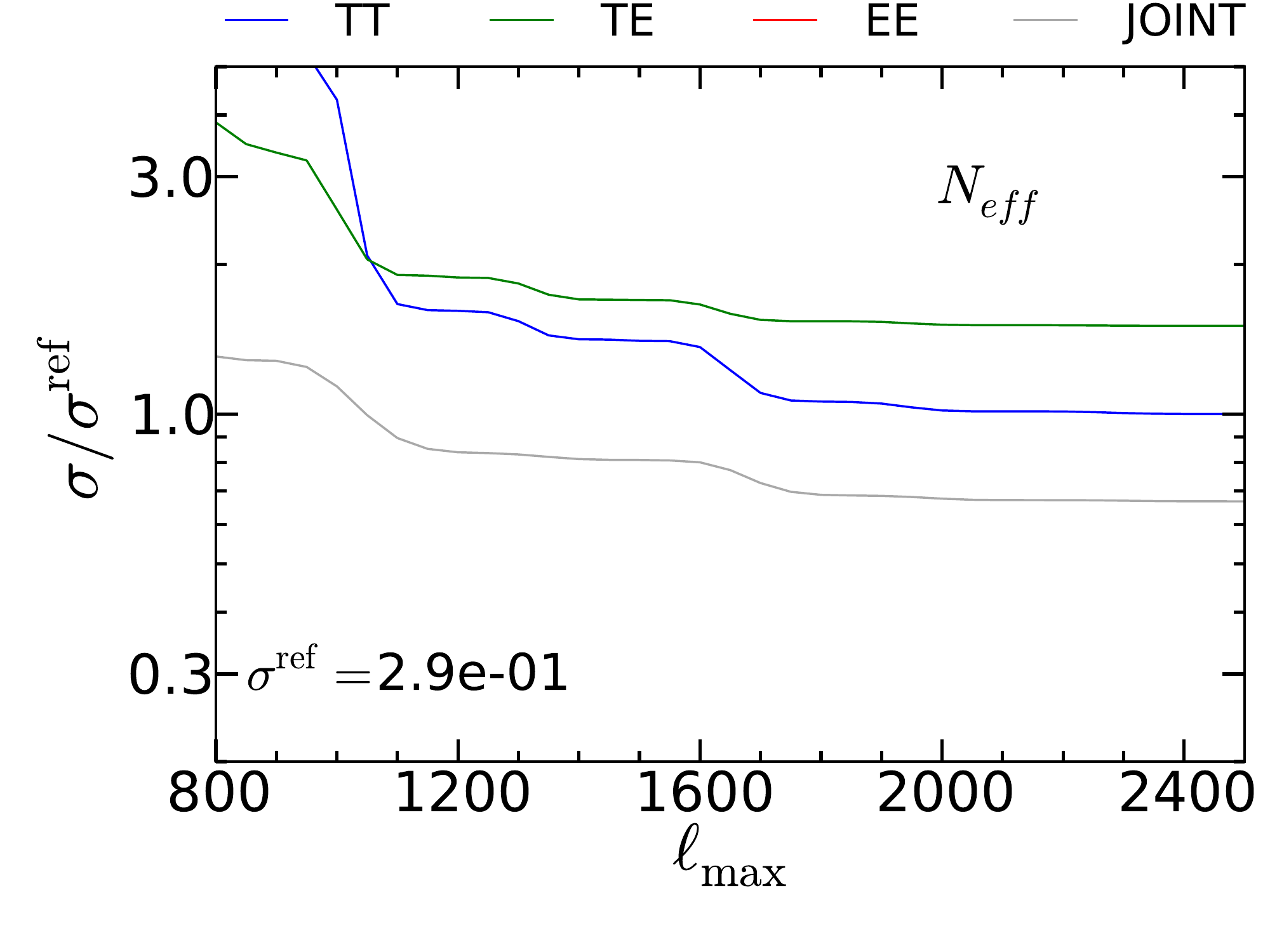}
\caption{Standard deviations on $N_{eff}$ as function of $\lmax$, normalized to the standard deviation $\sigma^{\rm ref}$ obtained from \TT\ with $\lmax=2500$. We consider a \CVL\ experiment (left) and a Planck-like full mission experiment (right) with $\lmin=30$ and a prior on $\tau$.}
\label{neff}
\end{figure*}

\begin{figure*}[t]
\centering
\includegraphics[angle=0,width=0.35\textwidth]{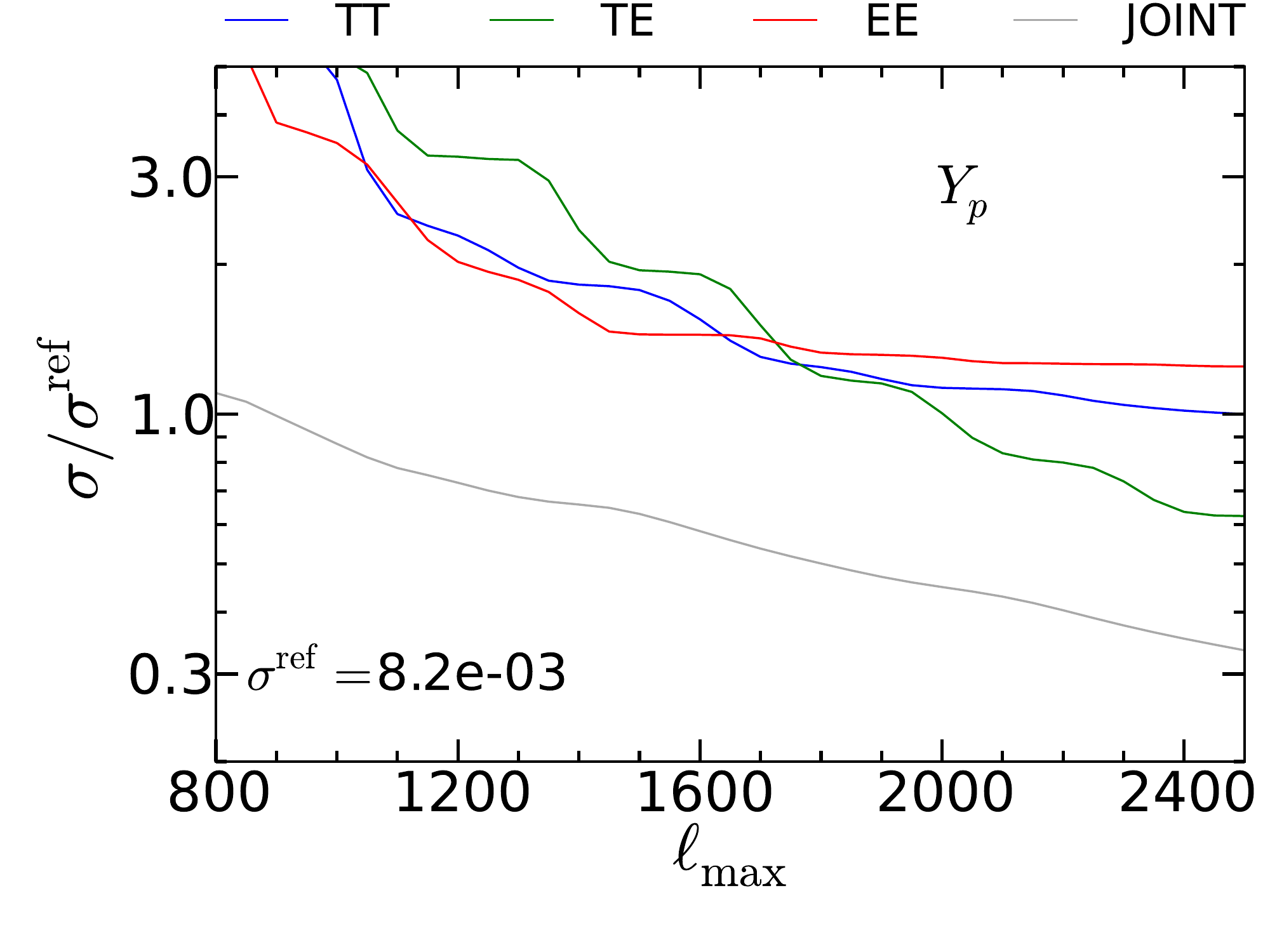}
\includegraphics[angle=0,width=0.35\textwidth]{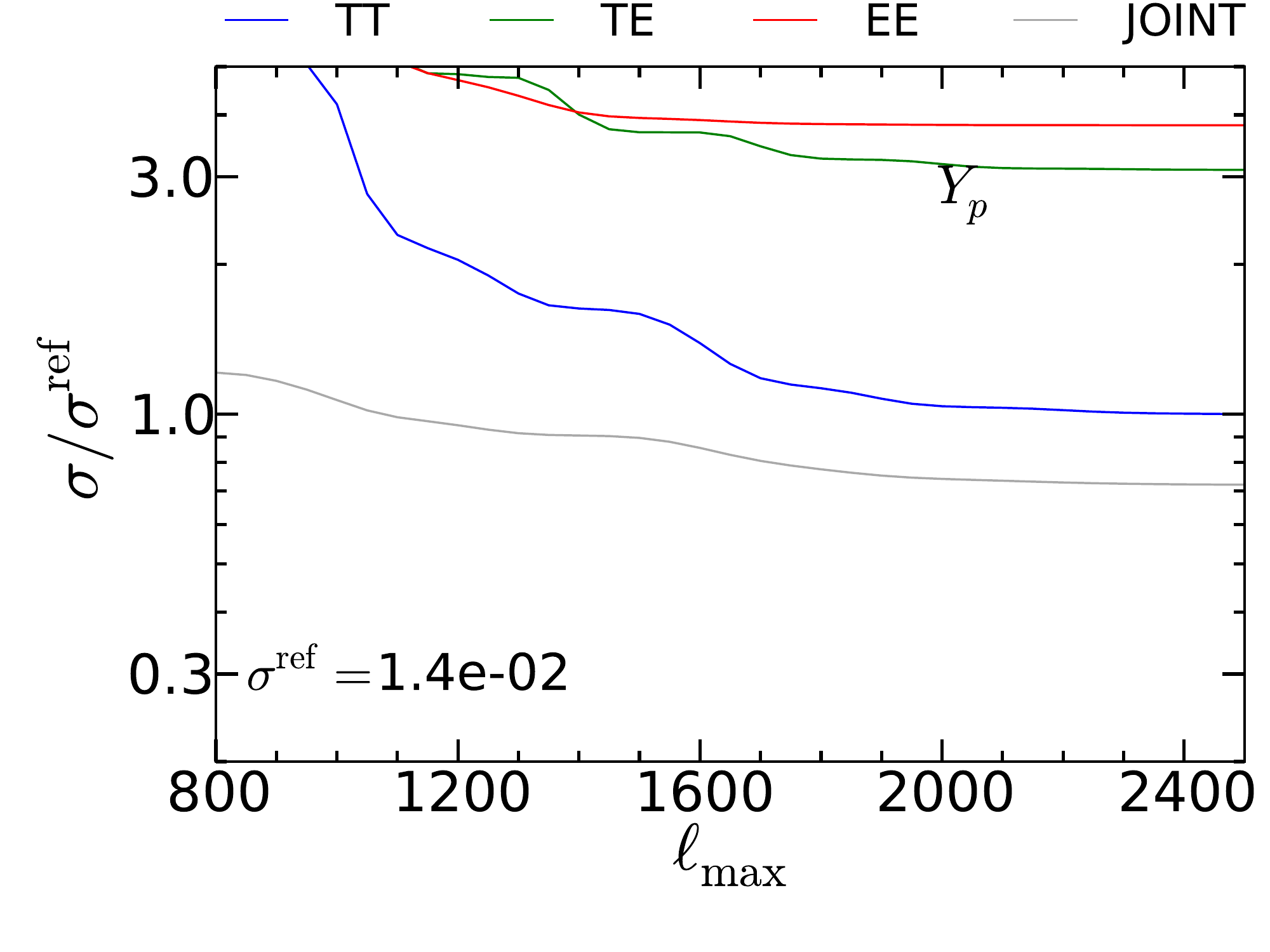}
\caption{Standard deviations on  $Y_p$  as function of $\lmax$, normalized to the standard deviation $\sigma^{\rm ref}$ obtained from \TT\ with $\lmax=2500$. We consider a \CVL\ experiment (left) and a Planck-like full mission experiment (right) with $\lmin=30$ and a prior on $\tau$.}
\label{helium}
\end{figure*}

\begin{figure*}[t]
\centering
\includegraphics[angle=0,width=0.35\textwidth]{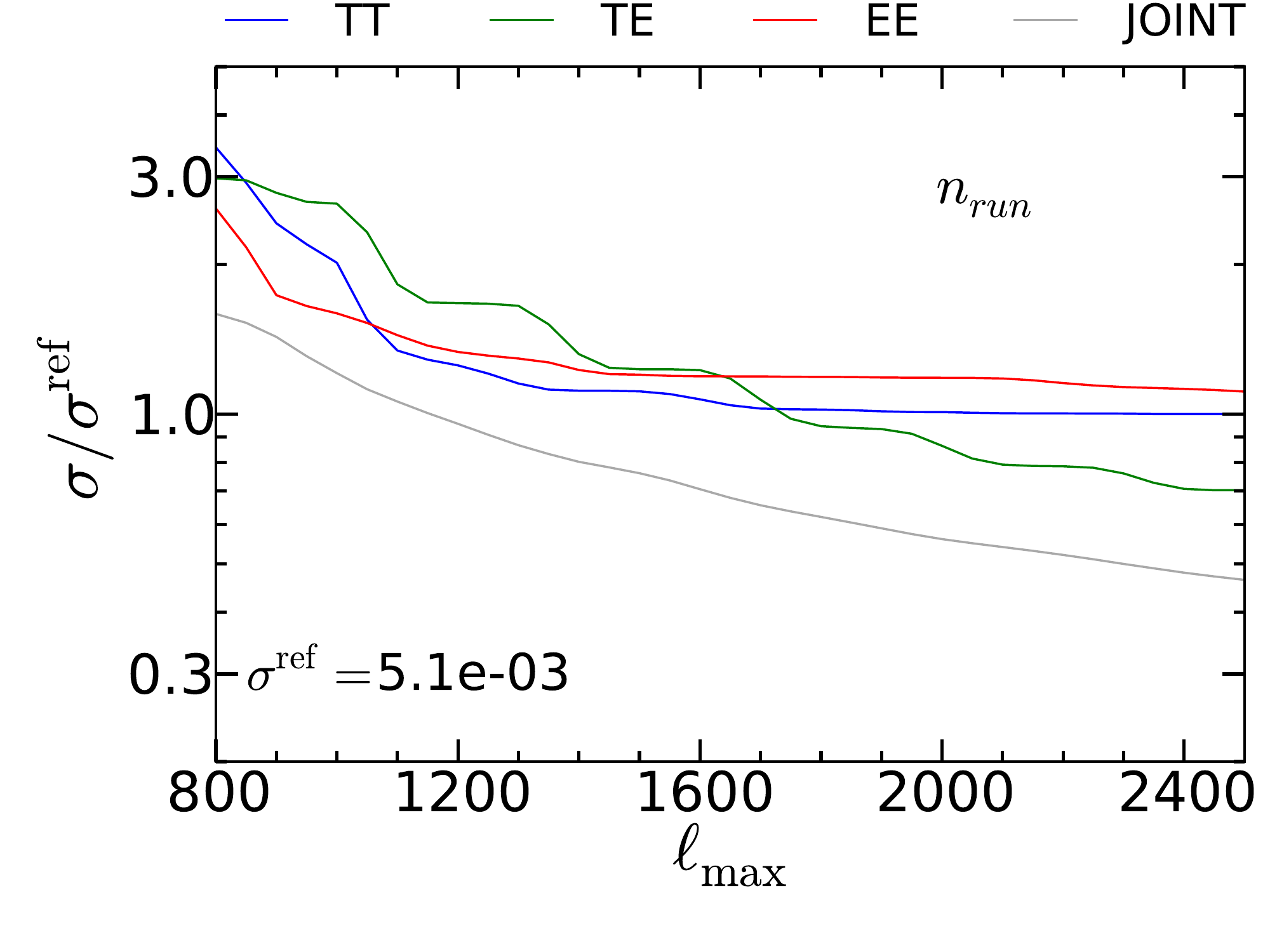}
\includegraphics[angle=0,width=0.35\textwidth]{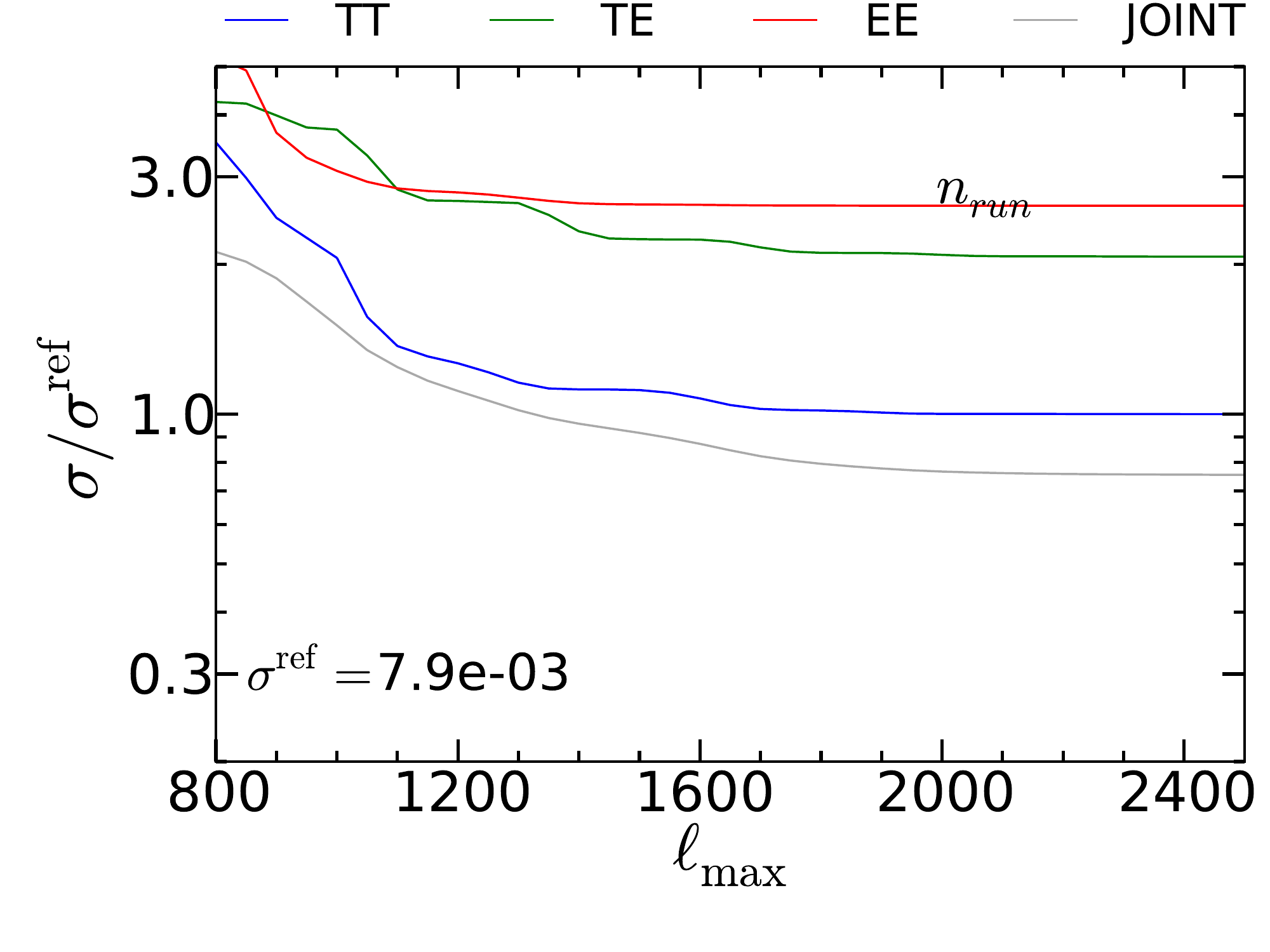}

\caption{Standard deviations on  $n_{run}$  as function of $\lmax$, normalized to the standard deviation $\sigma^{\rm ref}$ obtained from \TT\ with $\lmax=2500$. We consider a \CVL\ experiment (left) and a Planck-like full mission experiment (right) with $\lmin=30$ and a prior on $\tau$.}

\label{nrun}
\end{figure*}

 Figs. \ref{mnu}-\ref{nrun} show how the constraints on these parameters evolve with $\lmax$ for a \CVL\  experiment or a Planck-like full mission experiment. Contrary to what was previously found in the case of $\Lambda$CDM for a CVL experiment, only the neutrino mass $\sum m_\nu$ is better constrained by  \EE\ rather than  \TE\ at $\lmax=2500$, through the larger effect of lensing on  \EE. $N_{eff}$, $Y_p$ and $n_{run}$ are instead better constrained by  \TE. 
For a Planck-like experiment, on the other hand, we forecast that \TT\  is always superior in constraining all the extensions of $\Lambda$CDM. The case where the constraining power of  \TE\ seems to be the most interesting is $\neff$, for which we forecast $\sigma_{TE}(\neff)=0.43$ [14\%], to be compared to the constraint from  \TT, $\sigma_{TT}(\neff)=0.29$ [10\%], i.e. the constrain from  \TE\ is only a factor $1.5$ weaker.

\section{Conclusions}
\label{sec:conclusions}
We have forecasted the power to constrain cosmological parameters from the  \TT,  \EE\ or  \TE\ CMB power spectra taken separately. We find that for a cosmic variance limited experiment,  \EE\ is the best at constraining all cosmological parameters in a $\Lambda$CDM model compared to  \TE\ and  \TT\  alone. This fact holds true when different amount of information about cosmic reionization is included in the analysis, e.g. whether the polarized power spectra are cut at $\lmin=2$ or $\lmin=30$, or whether a prior on the reionization optical depth $\tau$ is included in the analysis. We find that most of the degeneracy breaking power of the  \EE\ power spectrum comes from $\ell$ ranges between $100-200$, a region where we expect the galactic dust contamination to be harder to clean. 
We have also shown for the first time the constraining power of the temperature-polarization cross-correlation power spectrum  \TE, finding that for a \CVL\ experiment it would also constrain parameters more efficiently than  \TT\  alone. The advantage of  \TE\ over  \EE\ however, is that the constraints are less affected when $\ell\lesssim 200$ are excluded from  the analysis, making it less sensitive to large-scale foreground modelling.
We have also shown that  \EE\ is the  best at constraining $\tau$, compared to  \TE\ or  \TT, either from the large scale polarization reionization bump, or from the degeneracy breaking effect of lensing at small scales. We find that observing the polarization bump provides constraints that are a factor $\sim 3$ stronger than the ones obtainable by the effect of lensing.

These findings are particularly interesting for future proposed CMB missions such as \textit{CORE} \cite{core} or \textit{PRISM} \cite{prism}, as such experiments are designed to be cosmic variance limited both in temperature and in polarization in wide multipole ranges. 

We also forecast constraints for a Planck-like satellite experiment.  
In spite of the much higher noise of polarization, we find that  \TE\ determines the dark matter density $\Omega_ch^2$  by $15\%$ better than  \TT, and the other parameters at a similar level of precision.
This result has never been forecasted before, and opens the possibility to verify the Planck results from  \TT\  with those from  \TE. The advantage of this procedure is that the two spectra are expected to have different dependencies on systematics and foregrounds. 
We emphasize here that our forecasts do not include marginalization over foreground parameters. However, the level of foreground contamination at small scales in polarization is expected to be lower than the one in temperature, so we expect that our broad conclusions  about the superiority of polarization should not depend on this factor.

\acknowledgments
This work was supported by Benjamin D. Wandelt's ANR Chaire d'Excellence ANR-10-CEXC-004-01. SG would like to thank Joe Silk, Aurelien Benoit-L{\'e}vy and Alessandro Melchiorri for comments on the manuscript.
\bibliography{TET_biblio}

\begin{thebibliography}{49}
\expandafter\ifx\csname natexlab\endcsname\relax\def\natexlab#1{#1}\fi
\expandafter\ifx\csname bibnamefont\endcsname\relax
  \def\bibnamefont#1{#1}\fi
\expandafter\ifx\csname bibfnamefont\endcsname\relax
  \def\bibfnamefont#1{#1}\fi
\expandafter\ifx\csname citenamefont\endcsname\relax
  \def\citenamefont#1{#1}\fi
\expandafter\ifx\csname url\endcsname\relax
  \def\url#1{\texttt{#1}}\fi
\expandafter\ifx\csname urlprefix\endcsname\relax\def\urlprefix{URL }\fi
\providecommand{\bibinfo}[2]{#2}
\providecommand{\eprint}[2][]{\url{#2}}

\bibitem[{\citenamefont{{Planck Collaboration}
  et~al.}(2013)\citenamefont{{Planck Collaboration}, {Ade}, {Aghanim},
  {Armitage-Caplan}, {Arnaud}, {Ashdown}, {Atrio-Barandela}, {Aumont},
  {Baccigalupi}, {Banday} et~al.}}]{planckcosmo}
\bibinfo{author}{\bibnamefont{{Planck Collaboration}}},
  \bibinfo{author}{\bibfnamefont{P.~A.~R.} \bibnamefont{{Ade}}},
  \bibinfo{author}{\bibfnamefont{N.}~\bibnamefont{{Aghanim}}},
  \bibinfo{author}{\bibfnamefont{C.}~\bibnamefont{{Armitage-Caplan}}},
  \bibinfo{author}{\bibfnamefont{M.}~\bibnamefont{{Arnaud}}},
  \bibinfo{author}{\bibfnamefont{M.}~\bibnamefont{{Ashdown}}},
  \bibinfo{author}{\bibfnamefont{F.}~\bibnamefont{{Atrio-Barandela}}},
  \bibinfo{author}{\bibfnamefont{J.}~\bibnamefont{{Aumont}}},
  \bibinfo{author}{\bibfnamefont{C.}~\bibnamefont{{Baccigalupi}}},
  \bibinfo{author}{\bibfnamefont{A.~J.} \bibnamefont{{Banday}}},
  \bibnamefont{et~al.}, \bibinfo{journal}{ArXiv e-prints}
  (\bibinfo{year}{2013}), \eprint{1303.5076}.

\bibitem[{\citenamefont{{Zaldarriaga}}(1997)}]{zaldarriaga97}
\bibinfo{author}{\bibfnamefont{M.}~\bibnamefont{{Zaldarriaga}}},
  \bibinfo{journal}{\prd} \textbf{\bibinfo{volume}{55}}, \bibinfo{pages}{1822}
  (\bibinfo{year}{1997}), \eprint{astro-ph/9608050}.

\bibitem[{\citenamefont{{Mortonson} and {Hu}}(2008)}]{mortonson08}
\bibinfo{author}{\bibfnamefont{M.~J.} \bibnamefont{{Mortonson}}}
  \bibnamefont{and} \bibinfo{author}{\bibfnamefont{W.}~\bibnamefont{{Hu}}},
  \bibinfo{journal}{\apj} \textbf{\bibinfo{volume}{672}}, \bibinfo{pages}{737}
  (\bibinfo{year}{2008}), \eprint{0705.1132}.

\bibitem[{\citenamefont{{Seljak} and {Zaldarriaga}}(1997)}]{seljak97}
\bibinfo{author}{\bibfnamefont{U.}~\bibnamefont{{Seljak}}} \bibnamefont{and}
  \bibinfo{author}{\bibfnamefont{M.}~\bibnamefont{{Zaldarriaga}}},
  \bibinfo{journal}{Physical Review Letters} \textbf{\bibinfo{volume}{78}},
  \bibinfo{pages}{2054} (\bibinfo{year}{1997}), \eprint{astro-ph/9609169}.

\bibitem[{\citenamefont{{Zaldarriaga} et~al.}(1997)\citenamefont{{Zaldarriaga},
  {Spergel}, and {Seljak}}}]{zaldarriagaspergel97}
\bibinfo{author}{\bibfnamefont{M.}~\bibnamefont{{Zaldarriaga}}},
  \bibinfo{author}{\bibfnamefont{D.~N.} \bibnamefont{{Spergel}}},
  \bibnamefont{and} \bibinfo{author}{\bibfnamefont{U.}~\bibnamefont{{Seljak}}},
  \bibinfo{journal}{\apj} \textbf{\bibinfo{volume}{488}}, \bibinfo{pages}{1}
  (\bibinfo{year}{1997}), \eprint{astro-ph/9702157}.

\bibitem[{\citenamefont{{Baumann} et~al.}(2009)\citenamefont{{Baumann},
  {Jackson}, {Adshead}, {Amblard}, {Ashoorioon}, {Bartolo}, {Bean},
  {Beltr{\'a}n}, {de Bernardis}, {Bird} et~al.}}]{baumann09}
\bibinfo{author}{\bibfnamefont{D.}~\bibnamefont{{Baumann}}},
  \bibinfo{author}{\bibfnamefont{M.~G.} \bibnamefont{{Jackson}}},
  \bibinfo{author}{\bibfnamefont{P.}~\bibnamefont{{Adshead}}},
  \bibinfo{author}{\bibfnamefont{A.}~\bibnamefont{{Amblard}}},
  \bibinfo{author}{\bibfnamefont{A.}~\bibnamefont{{Ashoorioon}}},
  \bibinfo{author}{\bibfnamefont{N.}~\bibnamefont{{Bartolo}}},
  \bibinfo{author}{\bibfnamefont{R.}~\bibnamefont{{Bean}}},
  \bibinfo{author}{\bibfnamefont{M.}~\bibnamefont{{Beltr{\'a}n}}},
  \bibinfo{author}{\bibfnamefont{F.}~\bibnamefont{{de Bernardis}}},
  \bibinfo{author}{\bibfnamefont{S.}~\bibnamefont{{Bird}}},
  \bibnamefont{et~al.}, in \emph{\bibinfo{booktitle}{American Institute of
  Physics Conference Series}}, edited by
  \bibinfo{editor}{\bibfnamefont{S.}~\bibnamefont{{Dodelson}}},
  \bibinfo{editor}{\bibfnamefont{D.}~\bibnamefont{{Baumann}}},
  \bibinfo{editor}{\bibfnamefont{A.}~\bibnamefont{{Cooray}}},
  \bibinfo{editor}{\bibfnamefont{J.}~\bibnamefont{{Dunkley}}},
  \bibinfo{editor}{\bibfnamefont{A.}~\bibnamefont{{Fraisse}}},
  \bibinfo{editor}{\bibfnamefont{M.~G.} \bibnamefont{{Jackson}}},
  \bibinfo{editor}{\bibfnamefont{A.}~\bibnamefont{{Kogut}}},
  \bibinfo{editor}{\bibfnamefont{L.}~\bibnamefont{{Krauss}}},
  \bibinfo{editor}{\bibfnamefont{M.}~\bibnamefont{{Zaldarriaga}}},
  \bibnamefont{and} \bibinfo{editor}{\bibfnamefont{K.}~\bibnamefont{{Smith}}}
  (\bibinfo{year}{2009}), vol. \bibinfo{volume}{1141} of
  \emph{\bibinfo{series}{American Institute of Physics Conference Series}}, pp.
  \bibinfo{pages}{10--120}, \eprint{0811.3919}.

\bibitem[{\citenamefont{{Bock} et~al.}(2009)\citenamefont{{Bock}, {Aljabri},
  {Amblard}, {Baumann}, {Betoule}, {Chui}, {Colombo}, {Cooray}, {Crumb}, {Day}
  et~al.}}]{bock09}
\bibinfo{author}{\bibfnamefont{J.}~\bibnamefont{{Bock}}},
  \bibinfo{author}{\bibfnamefont{A.}~\bibnamefont{{Aljabri}}},
  \bibinfo{author}{\bibfnamefont{A.}~\bibnamefont{{Amblard}}},
  \bibinfo{author}{\bibfnamefont{D.}~\bibnamefont{{Baumann}}},
  \bibinfo{author}{\bibfnamefont{M.}~\bibnamefont{{Betoule}}},
  \bibinfo{author}{\bibfnamefont{T.}~\bibnamefont{{Chui}}},
  \bibinfo{author}{\bibfnamefont{L.}~\bibnamefont{{Colombo}}},
  \bibinfo{author}{\bibfnamefont{A.}~\bibnamefont{{Cooray}}},
  \bibinfo{author}{\bibfnamefont{D.}~\bibnamefont{{Crumb}}},
  \bibinfo{author}{\bibfnamefont{P.}~\bibnamefont{{Day}}},
  \bibnamefont{et~al.}, \bibinfo{journal}{ArXiv e-prints}
  (\bibinfo{year}{2009}), \eprint{0906.1188}.

\bibitem[{\citenamefont{{Kinney}}(1998)}]{kenney98}
\bibinfo{author}{\bibfnamefont{W.~H.} \bibnamefont{{Kinney}}},
  \bibinfo{journal}{\prd} \textbf{\bibinfo{volume}{58}}, \bibinfo{eid}{123506}
  (\bibinfo{year}{1998}), \eprint{astro-ph/9806259}.

\bibitem[{\citenamefont{{BICEP2 Collaboration}
  et~al.}(2014)\citenamefont{{BICEP2 Collaboration}, {Ade}, {Aikin}, {Barkats},
  {Benton}, {Bischoff}, {Bock}, {Brevik}, {Buder}, {Bullock} et~al.}}]{bicep}
\bibinfo{author}{\bibnamefont{{BICEP2 Collaboration}}},
  \bibinfo{author}{\bibfnamefont{P.~A.~R.} \bibnamefont{{Ade}}},
  \bibinfo{author}{\bibfnamefont{R.~W.} \bibnamefont{{Aikin}}},
  \bibinfo{author}{\bibfnamefont{D.}~\bibnamefont{{Barkats}}},
  \bibinfo{author}{\bibfnamefont{S.~J.} \bibnamefont{{Benton}}},
  \bibinfo{author}{\bibfnamefont{C.~A.} \bibnamefont{{Bischoff}}},
  \bibinfo{author}{\bibfnamefont{J.~J.} \bibnamefont{{Bock}}},
  \bibinfo{author}{\bibfnamefont{J.~A.} \bibnamefont{{Brevik}}},
  \bibinfo{author}{\bibfnamefont{I.}~\bibnamefont{{Buder}}},
  \bibinfo{author}{\bibfnamefont{E.}~\bibnamefont{{Bullock}}},
  \bibnamefont{et~al.}, \bibinfo{journal}{ArXiv e-prints}
  (\bibinfo{year}{2014}), \eprint{1403.3985}.

\bibitem[{\citenamefont{{Spergel} and {Zaldarriaga}}(1997)}]{spergel97}
\bibinfo{author}{\bibfnamefont{D.~N.} \bibnamefont{{Spergel}}}
  \bibnamefont{and}
  \bibinfo{author}{\bibfnamefont{M.}~\bibnamefont{{Zaldarriaga}}},
  \bibinfo{journal}{Physical Review Letters} \textbf{\bibinfo{volume}{79}},
  \bibinfo{pages}{2180} (\bibinfo{year}{1997}), \eprint{astro-ph/9705182}.

\bibitem[{\citenamefont{{Mortonson} et~al.}(2009)\citenamefont{{Mortonson},
  {Dvorkin}, {Peiris}, and {Hu}}}]{mortonson09}
\bibinfo{author}{\bibfnamefont{M.~J.} \bibnamefont{{Mortonson}}},
  \bibinfo{author}{\bibfnamefont{C.}~\bibnamefont{{Dvorkin}}},
  \bibinfo{author}{\bibfnamefont{H.~V.} \bibnamefont{{Peiris}}},
  \bibnamefont{and} \bibinfo{author}{\bibfnamefont{W.}~\bibnamefont{{Hu}}},
  \bibinfo{journal}{\prd} \textbf{\bibinfo{volume}{79}}, \bibinfo{eid}{103519}
  (\bibinfo{year}{2009}), \eprint{0903.4920}.

\bibitem[{\citenamefont{{Hu} et~al.}(1997)\citenamefont{{Hu}, {Spergel}, and
  {White}}}]{huspergel}
\bibinfo{author}{\bibfnamefont{W.}~\bibnamefont{{Hu}}},
  \bibinfo{author}{\bibfnamefont{D.~N.} \bibnamefont{{Spergel}}},
  \bibnamefont{and} \bibinfo{author}{\bibfnamefont{M.}~\bibnamefont{{White}}},
  \bibinfo{journal}{\prd} \textbf{\bibinfo{volume}{55}}, \bibinfo{pages}{3288}
  (\bibinfo{year}{1997}), \eprint{astro-ph/9605193}.

\bibitem[{\citenamefont{{Durrer} et~al.}(2001)\citenamefont{{Durrer}, {Kunz},
  and {Melchiorri}}}]{durrer01}
\bibinfo{author}{\bibfnamefont{R.}~\bibnamefont{{Durrer}}},
  \bibinfo{author}{\bibfnamefont{M.}~\bibnamefont{{Kunz}}}, \bibnamefont{and}
  \bibinfo{author}{\bibfnamefont{A.}~\bibnamefont{{Melchiorri}}},
  \bibinfo{journal}{\prd} \textbf{\bibinfo{volume}{63}}, \bibinfo{eid}{081301}
  (\bibinfo{year}{2001}), \eprint{astro-ph/0010633}.

\bibitem[{\citenamefont{{Peiris} et~al.}(2003)\citenamefont{{Peiris},
  {Komatsu}, {Verde}, {Spergel}, {Bennett}, {Halpern}, {Hinshaw}, {Jarosik},
  {Kogut}, {Limon} et~al.}}]{peiris03}
\bibinfo{author}{\bibfnamefont{H.~V.} \bibnamefont{{Peiris}}},
  \bibinfo{author}{\bibfnamefont{E.}~\bibnamefont{{Komatsu}}},
  \bibinfo{author}{\bibfnamefont{L.}~\bibnamefont{{Verde}}},
  \bibinfo{author}{\bibfnamefont{D.~N.} \bibnamefont{{Spergel}}},
  \bibinfo{author}{\bibfnamefont{C.~L.} \bibnamefont{{Bennett}}},
  \bibinfo{author}{\bibfnamefont{M.}~\bibnamefont{{Halpern}}},
  \bibinfo{author}{\bibfnamefont{G.}~\bibnamefont{{Hinshaw}}},
  \bibinfo{author}{\bibfnamefont{N.}~\bibnamefont{{Jarosik}}},
  \bibinfo{author}{\bibfnamefont{A.}~\bibnamefont{{Kogut}}},
  \bibinfo{author}{\bibfnamefont{M.}~\bibnamefont{{Limon}}},
  \bibnamefont{et~al.}, \bibinfo{journal}{\apjs}
  \textbf{\bibinfo{volume}{148}}, \bibinfo{pages}{213} (\bibinfo{year}{2003}),
  \eprint{astro-ph/0302225}.

\bibitem[{\citenamefont{{Rocha} et~al.}(2004)\citenamefont{{Rocha}, {Trotta},
  {Martins}, {Melchiorri}, {Avelino}, {Bean}, and {Viana}}}]{rocha03}
\bibinfo{author}{\bibfnamefont{G.}~\bibnamefont{{Rocha}}},
  \bibinfo{author}{\bibfnamefont{R.}~\bibnamefont{{Trotta}}},
  \bibinfo{author}{\bibfnamefont{C.~J.~A.~P.} \bibnamefont{{Martins}}},
  \bibinfo{author}{\bibfnamefont{A.}~\bibnamefont{{Melchiorri}}},
  \bibinfo{author}{\bibfnamefont{P.~P.} \bibnamefont{{Avelino}}},
  \bibinfo{author}{\bibfnamefont{R.}~\bibnamefont{{Bean}}}, \bibnamefont{and}
  \bibinfo{author}{\bibfnamefont{P.~T.~P.} \bibnamefont{{Viana}}},
  \bibinfo{journal}{\mnras} \textbf{\bibinfo{volume}{352}}, \bibinfo{pages}{20}
  (\bibinfo{year}{2004}), \eprint{astro-ph/0309211}.

\bibitem[{\citenamefont{{Colombo} et~al.}(2009)\citenamefont{{Colombo},
  {Pierpaoli}, and {Pritchard}}}]{colombo09}
\bibinfo{author}{\bibfnamefont{L.~P.~L.} \bibnamefont{{Colombo}}},
  \bibinfo{author}{\bibfnamefont{E.}~\bibnamefont{{Pierpaoli}}},
  \bibnamefont{and} \bibinfo{author}{\bibfnamefont{J.~R.}
  \bibnamefont{{Pritchard}}}, \bibinfo{journal}{\mnras}
  \textbf{\bibinfo{volume}{398}}, \bibinfo{pages}{1621} (\bibinfo{year}{2009}),
  \eprint{0811.2622}.

\bibitem[{\citenamefont{{Tegmark} et~al.}(2000)\citenamefont{{Tegmark},
  {Eisenstein}, {Hu}, and {de Oliveira-Costa}}}]{tegmark00}
\bibinfo{author}{\bibfnamefont{M.}~\bibnamefont{{Tegmark}}},
  \bibinfo{author}{\bibfnamefont{D.~J.} \bibnamefont{{Eisenstein}}},
  \bibinfo{author}{\bibfnamefont{W.}~\bibnamefont{{Hu}}}, \bibnamefont{and}
  \bibinfo{author}{\bibfnamefont{A.}~\bibnamefont{{de Oliveira-Costa}}},
  \bibinfo{journal}{\apj} \textbf{\bibinfo{volume}{530}}, \bibinfo{pages}{133}
  (\bibinfo{year}{2000}), \eprint{astro-ph/9905257}.

\bibitem[{\citenamefont{{Eisenstein} et~al.}(1999)\citenamefont{{Eisenstein},
  {Hu}, and {Tegmark}}}]{eisenstein99}
\bibinfo{author}{\bibfnamefont{D.~J.} \bibnamefont{{Eisenstein}}},
  \bibinfo{author}{\bibfnamefont{W.}~\bibnamefont{{Hu}}}, \bibnamefont{and}
  \bibinfo{author}{\bibfnamefont{M.}~\bibnamefont{{Tegmark}}},
  \bibinfo{journal}{\apj} \textbf{\bibinfo{volume}{518}}, \bibinfo{pages}{2}
  (\bibinfo{year}{1999}), \eprint{astro-ph/9807130}.

\bibitem[{\citenamefont{{Bucher} et~al.}(2001)\citenamefont{{Bucher},
  {Moodley}, and {Turok}}}]{bucher01}
\bibinfo{author}{\bibfnamefont{M.}~\bibnamefont{{Bucher}}},
  \bibinfo{author}{\bibfnamefont{K.}~\bibnamefont{{Moodley}}},
  \bibnamefont{and} \bibinfo{author}{\bibfnamefont{N.}~\bibnamefont{{Turok}}},
  \bibinfo{journal}{Physical Review Letters} \textbf{\bibinfo{volume}{87}},
  \bibinfo{eid}{191301} (\bibinfo{year}{2001}), \eprint{astro-ph/0012141}.

\bibitem[{\citenamefont{{Wu} et~al.}(2014)\citenamefont{{Wu}, {Errard},
  {Dvorkin}, {Kuo}, {Lee}, {McDonald}, {Slosar}, and {Zahn}}}]{wu14}
\bibinfo{author}{\bibfnamefont{W.~L.~K.} \bibnamefont{{Wu}}},
  \bibinfo{author}{\bibfnamefont{J.}~\bibnamefont{{Errard}}},
  \bibinfo{author}{\bibfnamefont{C.}~\bibnamefont{{Dvorkin}}},
  \bibinfo{author}{\bibfnamefont{C.~L.} \bibnamefont{{Kuo}}},
  \bibinfo{author}{\bibfnamefont{A.~T.} \bibnamefont{{Lee}}},
  \bibinfo{author}{\bibfnamefont{P.}~\bibnamefont{{McDonald}}},
  \bibinfo{author}{\bibfnamefont{A.}~\bibnamefont{{Slosar}}}, \bibnamefont{and}
  \bibinfo{author}{\bibfnamefont{O.}~\bibnamefont{{Zahn}}},
  \bibinfo{journal}{ArXiv e-prints}  (\bibinfo{year}{2014}),
  \eprint{1402.4108}.

\bibitem[{\citenamefont{{Galli} et~al.}(2010)\citenamefont{{Galli},
  {Martinelli}, {Melchiorri}, {Pagano}, {Sherwin}, and {Spergel}}}]{galli}
\bibinfo{author}{\bibfnamefont{S.}~\bibnamefont{{Galli}}},
  \bibinfo{author}{\bibfnamefont{M.}~\bibnamefont{{Martinelli}}},
  \bibinfo{author}{\bibfnamefont{A.}~\bibnamefont{{Melchiorri}}},
  \bibinfo{author}{\bibfnamefont{L.}~\bibnamefont{{Pagano}}},
  \bibinfo{author}{\bibfnamefont{B.~D.} \bibnamefont{{Sherwin}}},
  \bibnamefont{and} \bibinfo{author}{\bibfnamefont{D.~N.}
  \bibnamefont{{Spergel}}}, \bibinfo{journal}{\prd}
  \textbf{\bibinfo{volume}{82}}, \bibinfo{eid}{123504} (\bibinfo{year}{2010}),
  \eprint{1005.3808}.

\bibitem[{\citenamefont{Collaboration et~al.}(2013)\citenamefont{Collaboration,
  Ade, Aghanim, Armitage-Caplan, Arnaud, Ashdown, Atrio-Barandela, Aumont,
  Baccigalupi, Banday et~al.}}]{2013arXiv1303.5075P}
\bibinfo{author}{\bibfnamefont{P.}~\bibnamefont{Collaboration}},
  \bibinfo{author}{\bibfnamefont{P.~A.~R.} \bibnamefont{Ade}},
  \bibinfo{author}{\bibfnamefont{N.}~\bibnamefont{Aghanim}},
  \bibinfo{author}{\bibfnamefont{C.}~\bibnamefont{Armitage-Caplan}},
  \bibinfo{author}{\bibfnamefont{M.}~\bibnamefont{Arnaud}},
  \bibinfo{author}{\bibfnamefont{M.}~\bibnamefont{Ashdown}},
  \bibinfo{author}{\bibfnamefont{F.}~\bibnamefont{Atrio-Barandela}},
  \bibinfo{author}{\bibfnamefont{J.}~\bibnamefont{Aumont}},
  \bibinfo{author}{\bibfnamefont{C.}~\bibnamefont{Baccigalupi}},
  \bibinfo{author}{\bibfnamefont{A.~J.} \bibnamefont{Banday}},
  \bibnamefont{et~al.}, \bibinfo{journal}{arXiv.org} p. \bibinfo{pages}{5075}
  (\bibinfo{year}{2013}), \eprint{1303.5075}.

\bibitem[{\citenamefont{Dunkley et~al.}(2013)\citenamefont{Dunkley, Calabrese,
  Sievers, Addison, Battaglia, Battistelli, Bond, Das, Devlin, Dunner
  et~al.}}]{Dunkley:2013vt}
\bibinfo{author}{\bibfnamefont{J.}~\bibnamefont{Dunkley}},
  \bibinfo{author}{\bibfnamefont{E.}~\bibnamefont{Calabrese}},
  \bibinfo{author}{\bibfnamefont{J.}~\bibnamefont{Sievers}},
  \bibinfo{author}{\bibfnamefont{G.~E.} \bibnamefont{Addison}},
  \bibinfo{author}{\bibfnamefont{N.}~\bibnamefont{Battaglia}},
  \bibinfo{author}{\bibfnamefont{E.~S.} \bibnamefont{Battistelli}},
  \bibinfo{author}{\bibfnamefont{J.~R.} \bibnamefont{Bond}},
  \bibinfo{author}{\bibfnamefont{S.}~\bibnamefont{Das}},
  \bibinfo{author}{\bibfnamefont{M.~J.} \bibnamefont{Devlin}},
  \bibinfo{author}{\bibfnamefont{R.}~\bibnamefont{Dunner}},
  \bibnamefont{et~al.}, \bibinfo{journal}{arXiv.org}  (\bibinfo{year}{2013}),
  \eprint{1301.0776v2}.

\bibitem[{\citenamefont{Reichardt et~al.}(2011)\citenamefont{Reichardt, Shaw,
  Zahn, Aird, Benson, Bleem, Carlstrom, Chang, Cho, Crawford
  et~al.}}]{Reichardt:2011il}
\bibinfo{author}{\bibfnamefont{C.~L.} \bibnamefont{Reichardt}},
  \bibinfo{author}{\bibfnamefont{L.}~\bibnamefont{Shaw}},
  \bibinfo{author}{\bibfnamefont{O.}~\bibnamefont{Zahn}},
  \bibinfo{author}{\bibfnamefont{K.~A.} \bibnamefont{Aird}},
  \bibinfo{author}{\bibfnamefont{B.~A.} \bibnamefont{Benson}},
  \bibinfo{author}{\bibfnamefont{L.~E.} \bibnamefont{Bleem}},
  \bibinfo{author}{\bibfnamefont{J.~E.} \bibnamefont{Carlstrom}},
  \bibinfo{author}{\bibfnamefont{C.~L.} \bibnamefont{Chang}},
  \bibinfo{author}{\bibfnamefont{H.~M.} \bibnamefont{Cho}},
  \bibinfo{author}{\bibfnamefont{T.~M.} \bibnamefont{Crawford}},
  \bibnamefont{et~al.}, \bibinfo{journal}{arXiv.org}  (\bibinfo{year}{2011}),
  \eprint{1111.0932v2}.

\bibitem[{\citenamefont{{Seiffert} et~al.}(2007)\citenamefont{{Seiffert},
  {Borys}, {Scott}, and {Halpern}}}]{seiffert07}
\bibinfo{author}{\bibfnamefont{M.}~\bibnamefont{{Seiffert}}},
  \bibinfo{author}{\bibfnamefont{C.}~\bibnamefont{{Borys}}},
  \bibinfo{author}{\bibfnamefont{D.}~\bibnamefont{{Scott}}}, \bibnamefont{and}
  \bibinfo{author}{\bibfnamefont{M.}~\bibnamefont{{Halpern}}},
  \bibinfo{journal}{\mnras} \textbf{\bibinfo{volume}{374}},
  \bibinfo{pages}{409} (\bibinfo{year}{2007}), \eprint{astro-ph/0610485}.

\bibitem[{\citenamefont{{Tucci} et~al.}(2005)\citenamefont{{Tucci},
  {Mart{\'{\i}}nez-Gonz{\'a}lez}, {Vielva}, and {Delabrouille}}}]{tucci04}
\bibinfo{author}{\bibfnamefont{M.}~\bibnamefont{{Tucci}}},
  \bibinfo{author}{\bibfnamefont{E.}~\bibnamefont{{Mart{\'{\i}}nez-Gonz{\'a}lez}}},
  \bibinfo{author}{\bibfnamefont{P.}~\bibnamefont{{Vielva}}}, \bibnamefont{and}
  \bibinfo{author}{\bibfnamefont{J.}~\bibnamefont{{Delabrouille}}},
  \bibinfo{journal}{\mnras} \textbf{\bibinfo{volume}{360}},
  \bibinfo{pages}{935} (\bibinfo{year}{2005}), \eprint{astro-ph/0411567}.

\bibitem[{\citenamefont{{Tegmark} et~al.}(1997)\citenamefont{{Tegmark},
  {Taylor}, and {Heavens}}}]{tegmark97}
\bibinfo{author}{\bibfnamefont{M.}~\bibnamefont{{Tegmark}}},
  \bibinfo{author}{\bibfnamefont{A.~N.} \bibnamefont{{Taylor}}},
  \bibnamefont{and} \bibinfo{author}{\bibfnamefont{A.~F.}
  \bibnamefont{{Heavens}}}, \bibinfo{journal}{\apj}
  \textbf{\bibinfo{volume}{480}}, \bibinfo{pages}{22} (\bibinfo{year}{1997}),
  \eprint{astro-ph/9603021}.

\bibitem[{\citenamefont{{Jungman} et~al.}(1996)\citenamefont{{Jungman},
  {Kamionkowski}, {Kosowsky}, and {Spergel}}}]{Jungman96}
\bibinfo{author}{\bibfnamefont{G.}~\bibnamefont{{Jungman}}},
  \bibinfo{author}{\bibfnamefont{M.}~\bibnamefont{{Kamionkowski}}},
  \bibinfo{author}{\bibfnamefont{A.}~\bibnamefont{{Kosowsky}}},
  \bibnamefont{and} \bibinfo{author}{\bibfnamefont{D.~N.}
  \bibnamefont{{Spergel}}}, \bibinfo{journal}{\prd}
  \textbf{\bibinfo{volume}{54}}, \bibinfo{pages}{1332} (\bibinfo{year}{1996}),
  \eprint{astro-ph/9512139}.

\bibitem[{\citenamefont{Benoit-L{\'e}vy
  et~al.}(2012)\citenamefont{Benoit-L{\'e}vy, Smith, and
  Hu}}]{2012PhRvD..86l3008B}
\bibinfo{author}{\bibfnamefont{A.}~\bibnamefont{Benoit-L{\'e}vy}},
  \bibinfo{author}{\bibfnamefont{K.~M.} \bibnamefont{Smith}}, \bibnamefont{and}
  \bibinfo{author}{\bibfnamefont{W.}~\bibnamefont{Hu}},
  \bibinfo{journal}{Physical Review D} \textbf{\bibinfo{volume}{86}},
  \bibinfo{pages}{123008} (\bibinfo{year}{2012}).

\bibitem[{\citenamefont{{Fendt} and {Wandelt}}(2007)}]{picopaper}
\bibinfo{author}{\bibfnamefont{W.~A.} \bibnamefont{{Fendt}}} \bibnamefont{and}
  \bibinfo{author}{\bibfnamefont{B.~D.} \bibnamefont{{Wandelt}}},
  \bibinfo{journal}{\apj} \textbf{\bibinfo{volume}{654}}, \bibinfo{pages}{2}
  (\bibinfo{year}{2007}), \eprint{astro-ph/0606709}.

\bibitem[{\citenamefont{{Knox}}(1995)}]{knox95}
\bibinfo{author}{\bibfnamefont{L.}~\bibnamefont{{Knox}}},
  \bibinfo{journal}{\prd} \textbf{\bibinfo{volume}{52}}, \bibinfo{pages}{4307}
  (\bibinfo{year}{1995}), \eprint{astro-ph/9504054}.

\bibitem[{\citenamefont{{Planck Collaboration}}(2005)}]{blubook}
\bibinfo{author}{\bibnamefont{{Planck Collaboration}}} (\bibinfo{year}{2005}),
  \bibinfo{note}{http://www.rssd.esa.int/
  SA/PLANCK/docs/Bluebook-ESA-SCI\%282005\%291\_V2.pdf}.

\bibitem[{\citenamefont{{Hinshaw} et~al.}(2013)\citenamefont{{Hinshaw},
  {Larson}, {Komatsu}, {Spergel}, {Bennett}, {Dunkley}, {Nolta}, {Halpern},
  {Hill}, {Odegard} et~al.}}]{wmap9}
\bibinfo{author}{\bibfnamefont{G.}~\bibnamefont{{Hinshaw}}},
  \bibinfo{author}{\bibfnamefont{D.}~\bibnamefont{{Larson}}},
  \bibinfo{author}{\bibfnamefont{E.}~\bibnamefont{{Komatsu}}},
  \bibinfo{author}{\bibfnamefont{D.~N.} \bibnamefont{{Spergel}}},
  \bibinfo{author}{\bibfnamefont{C.~L.} \bibnamefont{{Bennett}}},
  \bibinfo{author}{\bibfnamefont{J.}~\bibnamefont{{Dunkley}}},
  \bibinfo{author}{\bibfnamefont{M.~R.} \bibnamefont{{Nolta}}},
  \bibinfo{author}{\bibfnamefont{M.}~\bibnamefont{{Halpern}}},
  \bibinfo{author}{\bibfnamefont{R.~S.} \bibnamefont{{Hill}}},
  \bibinfo{author}{\bibfnamefont{N.}~\bibnamefont{{Odegard}}},
  \bibnamefont{et~al.}, \bibinfo{journal}{\apjs}
  \textbf{\bibinfo{volume}{208}}, \bibinfo{eid}{19} (\bibinfo{year}{2013}),
  \eprint{1212.5226}.

\bibitem[{\citenamefont{{Zaldarriaga}}(2004)}]{zaldarriaga04}
\bibinfo{author}{\bibfnamefont{M.}~\bibnamefont{{Zaldarriaga}}},
  \bibinfo{journal}{Measuring and Modeling the Universe} p.
  \bibinfo{pages}{309} (\bibinfo{year}{2004}), \eprint{astro-ph/0305272}.

\bibitem[{\citenamefont{{Holder} et~al.}(2003)\citenamefont{{Holder}, {Haiman},
  {Kaplinghat}, and {Knox}}}]{holder03}
\bibinfo{author}{\bibfnamefont{G.~P.} \bibnamefont{{Holder}}},
  \bibinfo{author}{\bibfnamefont{Z.}~\bibnamefont{{Haiman}}},
  \bibinfo{author}{\bibfnamefont{M.}~\bibnamefont{{Kaplinghat}}},
  \bibnamefont{and} \bibinfo{author}{\bibfnamefont{L.}~\bibnamefont{{Knox}}},
  \bibinfo{journal}{\apj} \textbf{\bibinfo{volume}{595}}, \bibinfo{pages}{13}
  (\bibinfo{year}{2003}), \eprint{astro-ph/0302404}.

\bibitem[{\citenamefont{{Bond} and {Efstathiou}}(1987)}]{bond87}
\bibinfo{author}{\bibfnamefont{J.~R.} \bibnamefont{{Bond}}} \bibnamefont{and}
  \bibinfo{author}{\bibfnamefont{G.}~\bibnamefont{{Efstathiou}}},
  \bibinfo{journal}{\mnras} \textbf{\bibinfo{volume}{226}},
  \bibinfo{pages}{655} (\bibinfo{year}{1987}).

\bibitem[{\citenamefont{{Zaldarriaga} and {Seljak}}(1998)}]{zaldarriaga98}
\bibinfo{author}{\bibfnamefont{M.}~\bibnamefont{{Zaldarriaga}}}
  \bibnamefont{and} \bibinfo{author}{\bibfnamefont{U.}~\bibnamefont{{Seljak}}},
  \bibinfo{journal}{\prd} \textbf{\bibinfo{volume}{58}}, \bibinfo{eid}{023003}
  (\bibinfo{year}{1998}), \eprint{astro-ph/9803150}.

\bibitem[{\citenamefont{{Hu} and {Sugiyama}}(1995)}]{hu95}
\bibinfo{author}{\bibfnamefont{W.}~\bibnamefont{{Hu}}} \bibnamefont{and}
  \bibinfo{author}{\bibfnamefont{N.}~\bibnamefont{{Sugiyama}}},
  \bibinfo{journal}{\apj} \textbf{\bibinfo{volume}{444}}, \bibinfo{pages}{489}
  (\bibinfo{year}{1995}), \eprint{arXiv:astro-ph/9407093}.

\bibitem[{\citenamefont{{Hu}}(1995)}]{huthesis}
\bibinfo{author}{\bibfnamefont{W.}~\bibnamefont{{Hu}}}, \bibinfo{journal}{ArXiv
  Astrophysics e-prints}  (\bibinfo{year}{1995}), \eprint{astro-ph/9508126}.

\bibitem[{\citenamefont{{Silk}}(1968)}]{silk68}
\bibinfo{author}{\bibfnamefont{J.}~\bibnamefont{{Silk}}},
  \bibinfo{journal}{\apj} \textbf{\bibinfo{volume}{151}}, \bibinfo{pages}{459}
  (\bibinfo{year}{1968}).

\bibitem[{\citenamefont{{Kaiser}}(1983)}]{kaiser83}
\bibinfo{author}{\bibfnamefont{N.}~\bibnamefont{{Kaiser}}},
  \bibinfo{journal}{\mnras} \textbf{\bibinfo{volume}{202}},
  \bibinfo{pages}{1169} (\bibinfo{year}{1983}).

\bibitem[{\citenamefont{{Kaplinghat} et~al.}(2003)\citenamefont{{Kaplinghat},
  {Chu}, {Haiman}, {Holder}, {Knox}, and {Skordis}}}]{Kaplinghat03}
\bibinfo{author}{\bibfnamefont{M.}~\bibnamefont{{Kaplinghat}}},
  \bibinfo{author}{\bibfnamefont{M.}~\bibnamefont{{Chu}}},
  \bibinfo{author}{\bibfnamefont{Z.}~\bibnamefont{{Haiman}}},
  \bibinfo{author}{\bibfnamefont{G.~P.} \bibnamefont{{Holder}}},
  \bibinfo{author}{\bibfnamefont{L.}~\bibnamefont{{Knox}}}, \bibnamefont{and}
  \bibinfo{author}{\bibfnamefont{C.}~\bibnamefont{{Skordis}}},
  \bibinfo{journal}{\apj} \textbf{\bibinfo{volume}{583}}, \bibinfo{pages}{24}
  (\bibinfo{year}{2003}), \eprint{astro-ph/0207591}.

\bibitem[{\citenamefont{{Hou} et~al.}(2014)\citenamefont{{Hou}, {Reichardt},
  {Story}, {Follin}, {Keisler}, {Aird}, {Benson}, {Bleem}, {Carlstrom}, {Chang}
  et~al.}}]{hou14}
\bibinfo{author}{\bibfnamefont{Z.}~\bibnamefont{{Hou}}},
  \bibinfo{author}{\bibfnamefont{C.~L.} \bibnamefont{{Reichardt}}},
  \bibinfo{author}{\bibfnamefont{K.~T.} \bibnamefont{{Story}}},
  \bibinfo{author}{\bibfnamefont{B.}~\bibnamefont{{Follin}}},
  \bibinfo{author}{\bibfnamefont{R.}~\bibnamefont{{Keisler}}},
  \bibinfo{author}{\bibfnamefont{K.~A.} \bibnamefont{{Aird}}},
  \bibinfo{author}{\bibfnamefont{B.~A.} \bibnamefont{{Benson}}},
  \bibinfo{author}{\bibfnamefont{L.~E.} \bibnamefont{{Bleem}}},
  \bibinfo{author}{\bibfnamefont{J.~E.} \bibnamefont{{Carlstrom}}},
  \bibinfo{author}{\bibfnamefont{C.~L.} \bibnamefont{{Chang}}},
  \bibnamefont{et~al.}, \bibinfo{journal}{\apj} \textbf{\bibinfo{volume}{782}},
  \bibinfo{eid}{74} (\bibinfo{year}{2014}), \eprint{1212.6267}.

\bibitem[{\citenamefont{Komatsu et~al.}(2011)}]{WMAP7}
\bibinfo{author}{\bibfnamefont{E.}~\bibnamefont{Komatsu}} \bibnamefont{et~al.}
  (\bibinfo{collaboration}{WMAP Collaboration}), \bibinfo{journal}{Astrophys.
  J. Suppl.} \textbf{\bibinfo{volume}{{192}}}, \bibinfo{pages}{18}
  (\bibinfo{year}{2011}), \eprint{1001.4538}.

\bibitem[{\citenamefont{Hou et~al.}(2011)\citenamefont{Hou, Keisler, Knox,
  Millea, and Reichardt}}]{hou11}
\bibinfo{author}{\bibfnamefont{Z.}~\bibnamefont{Hou}},
  \bibinfo{author}{\bibfnamefont{R.}~\bibnamefont{Keisler}},
  \bibinfo{author}{\bibfnamefont{L.}~\bibnamefont{Knox}},
  \bibinfo{author}{\bibfnamefont{M.}~\bibnamefont{Millea}}, \bibnamefont{and}
  \bibinfo{author}{\bibfnamefont{C.}~\bibnamefont{Reichardt}}
  (\bibinfo{year}{2011}), \eprint{1104.2333}.

\bibitem[{\citenamefont{{Bashinsky} and {Seljak}}(2004)}]{bashinsky04}
\bibinfo{author}{\bibfnamefont{S.}~\bibnamefont{{Bashinsky}}} \bibnamefont{and}
  \bibinfo{author}{\bibfnamefont{U.}~\bibnamefont{{Seljak}}},
  \bibinfo{journal}{\prd} \textbf{\bibinfo{volume}{69}}, \bibinfo{eid}{083002}
  (\bibinfo{year}{2004}), \eprint{astro-ph/0310198}.

\bibitem[{\citenamefont{{Ichikawa} et~al.}(2008)\citenamefont{{Ichikawa},
  {Sekiguchi}, and {Takahashi}}}]{ichikawa08}
\bibinfo{author}{\bibfnamefont{K.}~\bibnamefont{{Ichikawa}}},
  \bibinfo{author}{\bibfnamefont{T.}~\bibnamefont{{Sekiguchi}}},
  \bibnamefont{and}
  \bibinfo{author}{\bibfnamefont{T.}~\bibnamefont{{Takahashi}}},
  \bibinfo{journal}{\prd} \textbf{\bibinfo{volume}{78}}, \bibinfo{eid}{043509}
  (\bibinfo{year}{2008}), \eprint{0712.4327}.

\bibitem[{\citenamefont{{The COrE Collaboration}
  et~al.}(2011)\citenamefont{{The COrE Collaboration}, {Armitage-Caplan},
  {Avillez}, {Barbosa}, {Banday}, {Bartolo}, {Battye}, {Bernard}, {de
  Bernardis}, {Basak} et~al.}}]{core}
\bibinfo{author}{\bibnamefont{{The COrE Collaboration}}},
  \bibinfo{author}{\bibfnamefont{C.}~\bibnamefont{{Armitage-Caplan}}},
  \bibinfo{author}{\bibfnamefont{M.}~\bibnamefont{{Avillez}}},
  \bibinfo{author}{\bibfnamefont{D.}~\bibnamefont{{Barbosa}}},
  \bibinfo{author}{\bibfnamefont{A.}~\bibnamefont{{Banday}}},
  \bibinfo{author}{\bibfnamefont{N.}~\bibnamefont{{Bartolo}}},
  \bibinfo{author}{\bibfnamefont{R.}~\bibnamefont{{Battye}}},
  \bibinfo{author}{\bibfnamefont{J.}~\bibnamefont{{Bernard}}},
  \bibinfo{author}{\bibfnamefont{P.}~\bibnamefont{{de Bernardis}}},
  \bibinfo{author}{\bibfnamefont{S.}~\bibnamefont{{Basak}}},
  \bibnamefont{et~al.}, \bibinfo{journal}{ArXiv e-prints}
  (\bibinfo{year}{2011}), \eprint{1102.2181}.

\bibitem[{\citenamefont{{PRISM Collaboration} et~al.}(2013)\citenamefont{{PRISM
  Collaboration}, {Andre}, {Baccigalupi}, {Barbosa}, {Bartlett}, {Bartolo},
  {Battistelli}, {Battye}, {Bendo}, {Bernard} et~al.}}]{prism}
\bibinfo{author}{\bibnamefont{{PRISM Collaboration}}},
  \bibinfo{author}{\bibfnamefont{P.}~\bibnamefont{{Andre}}},
  \bibinfo{author}{\bibfnamefont{C.}~\bibnamefont{{Baccigalupi}}},
  \bibinfo{author}{\bibfnamefont{D.}~\bibnamefont{{Barbosa}}},
  \bibinfo{author}{\bibfnamefont{J.}~\bibnamefont{{Bartlett}}},
  \bibinfo{author}{\bibfnamefont{N.}~\bibnamefont{{Bartolo}}},
  \bibinfo{author}{\bibfnamefont{E.}~\bibnamefont{{Battistelli}}},
  \bibinfo{author}{\bibfnamefont{R.}~\bibnamefont{{Battye}}},
  \bibinfo{author}{\bibfnamefont{G.}~\bibnamefont{{Bendo}}},
  \bibinfo{author}{\bibfnamefont{J.-P.} \bibnamefont{{Bernard}}},
  \bibnamefont{et~al.}, \bibinfo{journal}{ArXiv e-prints}
  (\bibinfo{year}{2013}), \eprint{1306.2259}.

\end{thebibliography}
\bibliographystyle{apsrev}

\end{document}